\documentclass[10pt,twocolumn]{article} 
\usepackage{simpleConference}
\usepackage{times}
\usepackage{graphicx}
\usepackage{amssymb}
\usepackage{url,hyperref}

\usepackage{caption}
\usepackage{subcaption}

\usepackage{xpatch}
\usepackage{amsthm}
\usepackage{makecell}
\usepackage{multirow}
\usepackage{pgfplots}
\usepackage[ruled,vlined,linesnumbered]{algorithm2e}
\usepackage{bbold}

\usepackage{wrapfig}
\newtheorem{theorem}{Theorem}
\usepackage{placeins} 
\usepackage[normalem]{ulem} 

\input{tex/tikzLib.tex} 
\usetikzlibrary{overlay-beamer-styles,positioning,backgrounds,calc,shadings,shapes.arrows,shapes.symbols,shapes.geometric,shadows,patterns}
\usepackage{pgf}
\usepgfplotslibrary{statistics}

\usepackage{float}

\usepackage[acronym]{glossaries}
\makeglossaries
\newacronym{tcp}{TCP}{Transmission Control Protocol}
\newacronym{dctcp}{DCTCP}{Data center Transmission Control Protocol}
\newacronym{dwtcp}{DWTCP}{Double Window Transmission Control Protocol}
\newacronym{rtt}{RTT}{Round Trip Time}
\newacronym{rto}{RTO}{Retransmission Timeout}
\newacronym{cdf}{CDF}{Cumulative Distribution Function}
\newacronym{cwnd}{CWND}{Congestion Window}
\newacronym{bdp}{BDP}{Bandwidth-Delay Product}

\newacronym{bbr}{BBR}{Bottleneck Bandwidth and Round-trip propagation time}
\newacronym{int}{INT}{Inband Network Telemetry}
\newacronym{ecn}{ECN}{Explicit Congestion Notification}
\newacronym{udp}{UDP}{User Datagram Protocol}
\newacronym{dre}{DRE}{Discounting Rate Estimator}
\newacronym{ecmp}{ECMP}{Equal-cost multi-path routing}
\newacronym{ode}{ODE}{Ordinary Differential Equation}

\newacronym[longplural={Graphics Processing Units}]{gpu}{GPU}{Graphics Processing Unit}

\newcommand{\proto}{DWTCP\,}
\newcommand{\service}{\texttt{Scout\,}}
\newcommand{\Service}{\texttt{Scout}}

\newcommand\codestyle[1]{\begingroup \ttfamily #1\endgroup}

\begin{document}

\title{\Large \bf DWTCP: Ultra low latency congestion control protocol for Data centers}

\author{Shiva Ketabi \\
Sepehr Abbasi \\
University of Toronto, Canada \\
\texttt{Shiva.Ketabi@huawei.com}  \\
\texttt{Sepehr.Abbasi@huawei.com}  \\
\and
Ali Munir \\
Mahmoud Bahnasy \\
Yashar Ganjali \\
Huawei Technologies Co., Ltd. \\
\texttt{ali.munir@huawei.com}  \\
\texttt{mahmoud.mohamed.bahnasy@huawei.com}  \\
\texttt{Yashar.Ganjali@huawei.com}  \\
}


\maketitle
\thispagestyle{empty}

\begin{abstract}
    Congestion control algorithms rely on a variety of congestion signals (packet loss, Explicit Congestion Notification, delay, etc.) to achieve fast convergence, high utilization, and fairness among flows.
    A key limitation of these congestion signals is that they are either late in feedback or they incur significant overheads.
    An ideal congestion control must discover any available bandwidth in the network, detect congestion as soon as link utilization approaches full capacity, and react timely to avoid queuing and packet drops, without significant overheads.
    To this end, this work proposes \service service that leverages priority queues to infer ``bandwidth availability'' and ``link busyness'' at the host.
    The key observation here is that as the high priority queue (HPQ) gets busier, the low priority queue (LPQ) is served less.
    Therefore, the state of the link can be observed from the LPQ and any congestion can be detected several RTTs earlier than observing the HPQ. 
    We propose a new transport protocol, Double-Window Transmission Control Protocol (DWTCP) that builds upon the \service service to dynamically adjust its congestion window.
    Our testbed and simulation-based evaluation demonstrates that \service enables a data center transport to achieve high throughput, near-zero queues, lower latency, and high fairness.
\end{abstract}

\section{Introduction}
Congestion Control (CC) algorithms rely on a variety of congestion signals (such as packet loss~\cite{allman1999tcp}, \gls{ecn}~\cite{alizadeh2010data,munir2013minimizing, munir2014friends}, delay~\cite{swift2020,Timely,TCPVegas,BBR}, or INT~\cite{inta}) to achieve fast convergence, short flow completion times, small queues, high link utilization, and high fairness among the flows.
Today's data centers serve applications that generate highly dynamic network traffic.
The performance of these applications is strictly dependent on the efficiency of the underlying CC algorithm.
The efficiency of CC is tightly coupled with how quickly and accurately it can detect congestion in the network.

Existing congestion signals have limitations that impact CC solutions relying on them.
Some congestion signals provide late feedback, i.e., they provide network state information when congestion has already happened in the network (e.g., loss~\cite{allman1999tcp} or delay~\cite{Timely}).
The fundamental reason for the late feedback is built-in to the inherent design of these congestion signals such that all these signals rely on the behavior or state of the queues in the network.
For example, packet drops happen when the queue is already full, and \gls{ecn} is marked when the queue is above a certain threshold. Similarly, delay is observed when the link is already over-provisioned.
Due to the limitations of these signaling mechanisms, the underlying congestion control algorithms use conservative control laws and struggle to achieve fast convergence, high link utilization and fairness.
Recent solutions such as \gls{int}~\cite{inta,intb} and credit-based~\cite{cho2017credit} provide rich signals that can alleviate this problem. However, these solutions incur overheads and require changes in the switches, which may not always be feasible. 

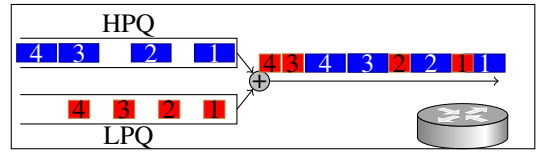
\begin{figure}[ht]
    \vspace{-0.1in}
	\centering
	\begin{tikzpicture} [scale=0.6]
		\def\qlen{4.8}
		\def\qheight{0.65}
		\def\qxshift{0.5}
		\def\qyshift{0.3}
		\node[scale=0.2] at (4.5, -1) {\router{}};
		\draw[draw=black] (-5.5,-1.5) rectangle ++(11.6,3.2);
		\node [circle,draw=black, fill=white!50!gray, inner sep=0pt,minimum size=7pt] (qserver) at (0,0) {+};
		\foreach \i in {1,-1}
		{
			\draw[black] (-\qlen-\qxshift, \i*\qyshift) -- (-\qxshift, \i*\qyshift) -- (-\qxshift, \i*\qyshift+\i*\qheight) -- (-\qlen-\qxshift, \i*\qyshift+\i*\qheight) node[pos=1, yshift=5] {}; 
		}
		\node[yshift=5] at (-0.5*\qlen-\qxshift, \qyshift+\qheight) {HPQ}; 
		\node[yshift=-5] at (-0.5*\qlen-\qxshift, -\qyshift-\qheight) {LPQ}; 
        
		\draw[black, ->] (-\qxshift, \qyshift+0.5*\qheight) -- (qserver); 
		\draw[black, ->] (-\qxshift, -\qyshift-0.5*\qheight) -- (qserver); 
		\draw[black, ->] (qserver) --  (\qxshift+\qlen, 0); 

		\newcounter{count} 
		\setcounter{count}{0}
		\foreach \i/\j/\l/\m in {-1/-1/5/4.5,-2.4/-2/3.8/3.1, -4/-3/2.4/0.74, -4.94/-4/1.459/.23} 
		{
			\stepcounter{count};
			\node[draw=blue!50!black, fill=blue,align=center, minimum width=15,text=white, inner sep=0cm] at (\i,\qyshift+0.5*\qheight) {\thecount}; 
			\node[draw=brown, fill=red,align=center, minimum width=8, inner sep=0cm] at (\j,-\qyshift-0.5*\qheight) {\thecount}; 
			\node[draw=blue!50!black, fill=blue,align=center, minimum width=15,text=white, inner sep=0cm] at (\l,0.4) {\thecount}; 
			\node[draw=brown, fill=red,align=center, minimum width=8, inner sep=0cm] at (\m,0.4) {\thecount}; 
		}
	\end{tikzpicture}
	\caption{
    Under strict priority queueing, low priority packets can only be served if the high priority queue is idle. This effect can be used as a congestion signal.}
	\label{FIG:scoutAffect}
    \vspace{-0.15in}
\end{figure}

In this work, through \service~service, we demonstrate that one can get an \textit{early} and \textit{amplified} congestion signal from the network switches, using a basic feature (priority queues) available in commodity switches, without the need for switch modification or imposing any complicated algorithm at the host.
\service~ service is built based on a simple yet effective idea of using priority queues in the switches to get the congestion state. 
The key observation here is that as the high priority queue (HPQ) gets busier (higher link utilization), the low priority queue (LPQ) gets less chance to be served (Figure~\ref{FIG:scoutAffect}). 
Therefore, the state of the link can be observed from the LPQ and any congestion can be detected several RTTs before observing the HPQ building up as done by the existing congestion signals.

\service service allows end-hosts to send \service traffic (with low priority) to the receiver alongside the normal traffic (with high priority) and then monitors the \service acknowledgments (ACKs) from the receiver to infer \textit{``bandwidth availability''} and \textit{``link busyness''} (or early indication of congestion).
At the switch side, as shown in Figure~\ref{FIG:scoutAffect}, the normal traffic (blue packets) is served with strict high priority, while \service~ traffic (red packets) is served in the low priority queue, i.e., only when the output port is idle.
In this setting, \service~ traffic can only fill in the gaps between normal traffic packets.
Such behavior is an implicit guarantee from the network switches that it has available bandwidth equal to the number of \service packets.
%
%
In addition, the absence (loss) of, or a delay in serving \service packets indicates a busy link which provides an early indication of congestion. 

\service service is fundamentally different from existing credit-based schemes~\cite{cho2017credit}, probing-based schemes~\cite{anderson2006pcp, eccp, bart, pathChirp, topp}, and INT-based schemes~\cite{li2019hpcc}.
%
%
Compared to credit based schemes~\cite{cho2017credit}, \service service does not need to set aside any bandwidth for processing credits at the switches; instead, it uses network bandwidth only when there is no data traffic.
Compared to existing probing based schemes~\cite{eccp, bart, pathChirp, topp}, \service service uses low priority queues to probe the network, which not only provides significant amplification in congestion signal, but also significantly reduces the overhead of \service traffic, as the \service traffic is served only if there is available bandwidth along the path. 
%
%
Compared to INT based schemes, \service service does not require switches to make any computations and update packet headers, and has a lower overhead.

To demonstrate the effectiveness of \Service, we present a new transport protocol, \gls{dwtcp}, that 
dynamically adjusts the congestion window size of a flow based on the \service feedback. 
%
%
The key idea of \proto is to increase aggressively if the network has more bandwidth available and slow down as the link busyness approaches the full link utilization. Since \service provides early and amplified congestion signal, this aggressiveness improves convergence as well as fairness. 
%

%
\proto is a complete end-host solution and requires no changes in the switch architecture. 
At its core, \proto uses transmission control protocol (TCP) as the flow control mechanism.
Bandwidth availability (from \Service) informs the sender of any available bandwidth in the network, and as a result, senders increase their window aggressively for faster convergence. 
Similarly, the loss and delay in \service ACKs inform the sender about the link busyness which is used by \proto to adjust its transmission rate. 
%
%

A \service based transport has the following key properties:
\begin{itemize}
	\setlength{\itemsep}{0pt plus 1pt}
    \item \service enables a congestion control algorithm to discover any available bandwidth in the network, detect congestion significantly earlier than existing congestion signals (i.e., as soon as the link utilization approaches full capacity), and to react timely to avoid queue buildup and packet drops.
	\item \service does not incur much overhead in terms of network bandwidth and switch buffer utilization. First, \service packets are served only if there is available bandwidth. Second, \service packets are very small (64 bytes in our implementation). Third, \service does not require switches to reserve too much buffer capacity for LPQ. We demonstrate that even LPQ of 500 bytes (i.e., 5-10, 64-byte \service packets) results in the best performance.
    \item \proto detects and reacts quickly to network changes and is able to maintain near-zero queues, converges fast (maintain similar throughput as DCTCP even with near-zero queues), and significantly reduces the latency of short flows (details in \S\ref{sec:motiv}).

	\item \proto is not sensitive to parameter settings. We mathematically analyze the stability of \proto and deduce the stability conditions (see Appendix~\ref{SEC:analysis}). Evaluation also suggests that the recommended parameter settings provide best performance across a range of dynamic network settings.
\end{itemize}

We evaluate \proto using a combination of NS3 simulations and a Linux kernel testbed implementation. 
For the latter, we implement a loadable kernel module that does not require any changes in Linux's TCP stack.
We compare against DCTCP~\cite{alizadeh2010data}, TCP Reno~\cite{allman1999tcp} and HPCC~\cite{li2019hpcc} using a range of workloads and network settings.
%
For the long-lived flows, \proto improves convergence time by 2x, and achieves near perfect fairness.
For more realistic data center settings, \proto maintains near-zero queue lengths without affecting throughput or incurring long flow completion times even at 80\% load.
Moreover, at high loads, compared to DCTCP (and HPCC), \proto improves the mean latency and tail latency by up to  1x (1.7x) and 3x (5x), respectively, for the datamining workload and up to 2x (3x) and 3.5x (5x), respectively, for the websearch workload. 
At low loads, HPCC performs 31\% better than the \proto.
We observe similar trends in our testbed results.
Finally, \service incurs very little overhead in the network
(less than 0.2\% in our experiments).

\section{Design Rationale} \label{sec:motiv}
Timely detection of congestion is crucial to the transport layer design. 
Traditionally, CC has relied on reactive congestion signals (such as packet loss~\cite{allman1999tcp}, delay~\cite{Timely,swift2020}, or ECN~\cite{alizadeh2010data}) that react to queue occupancy or the proactive congestion signals (such as credits or INT) that predict the link state even before congestion occurs.
Reactive congestion signals provide delayed feedback (due to their reliance on the queue length) when the congestion has already happened.
Proactive congestion signals provide early feedback, but current implementations incur significant overheads in terms of either probing (Credits~\cite{cho2017credit}) or adding extra header fields (XCP~\cite{katabi2002congestion}, INT~\cite{inta}, etc.).

Next, we show that one can get much early and amplified congestion signals, compared to packet loss and delay, by sending low rate and low priority \service packets and observing the impact of normal traffic (high priority) on \service packets.

\begin{figure}[htp]
    \centering
    \begin{subfigure}[b]{0.48\linewidth}
        \begin{tikzpicture} [scale=0.4, font=\LARGE]
            \begin{axis}[legend style={at={(0.2,.95)},anchor=north,legend cell align=left, fill opacity=0.6, text opacity=1},
                xlabel=Link Utilization of high priority Traffic (\%), 
                ylabel= Que. OWD Ratio,
                y label style={at={(0.06,0.5)}}, height=6.cm, width=9cm,
                ymajorgrids=true,xmajorgrids=true,]
                \addplot[color=black, mark=x, line width=2.0pt] table [x=u, y=r, col sep=comma] {figures/correlations/OWDRatio.csv};
            \end{axis}
        \end{tikzpicture}
        \caption{Low priority queue delay / high priority queue delay}
        \label{FIG:amplification_linkvsdelay}
    \end{subfigure}
    \hspace*{\fill}
    \begin{subfigure}[b]{0.48\linewidth}
        \centering
        \begin{tikzpicture} [scale=0.4, font=\LARGE]
            \begin{axis}[scatter/classes={a={mark=x,draw=black}},
                xlabel=Traffic Aggressivness (Mbps/s),
                ylabel=  Time $\Delta$ (RTT),
                y label style={at={(0.02,0.5)}},
                height=6.cm, width=9cm,
                ymajorgrids=true,xmajorgrids=true,
                ]
                \addplot[color=black, mark=o, line width=2.0pt] table [x=t, y=deltaR, col sep=comma] {figures/correlations/timeToCongestionVsRateChange.csv};
            \end{axis}
        \end{tikzpicture}
        \caption{Packet loss time difference between HPQ and LPQ}
        \label{FIG:amplification_timeToFullQueue2}
    \end{subfigure}
    \caption{Amplification effect of low priority Queues. a) Packets in low priority queue (LPQ) observe significant delays compared to the packets in the high priority queue (HPQ) as link capacity approaches maximum capapcity. 
        b) Packets in LPQ observe packet drop hundreds of RTTs before packets in the HPQ, depending on the aggressiveness of ingress traffic.} 
    \label{FIG:amplification}
    \vspace{-0.1in}
\end{figure}
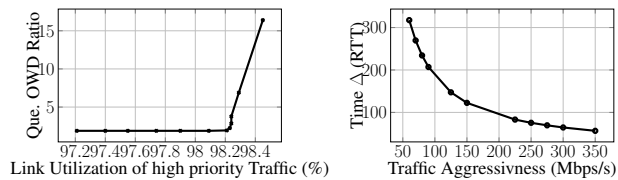

\vspace{-0.1in}
\subsection{Signal from Low Priority Queue }
Observing low priority queue (LPQ) serves as an early 
    (several RTTs)
indication of congestion and provides significant amplification in congestion signal (such as loss and delay). 
To understand, let us consider a single bottleneck topology, where a fixed number of senders generate UDP traffic with gradually increasing rates.
%
We compare the delay and packet loss experienced by the packets in the low priority queue and the high priority queue.

Figure~\ref{FIG:amplification_linkvsdelay} shows the amplification in one way delay (OWD) of low priority traffic compared to one way delay of high priority traffic.
We observe that as the link utilization approaches 100\%, the packets in low priority queue get fewer chances to be served and hence experience much higher delays. 
Figure~\ref{FIG:amplification_timeToFullQueue2} shows a significant difference in the time we observe packet loss from the low priority traffic compared to the high priority traffic.
The magnitude of these gains can vary based on the traffic arrival patterns and queue size of the high priority queue at the switches, but the amplification (higher delay) and early notification (loss observed sooner) are properties that can be used to enhance congestion control.
%

\vspace{-0.1in}
\subsection{Benefits for Congestion Control} 
\label{sec:benefits}
Early and amplified congestion feedback from the low priority queues help congestion control in terms of high throughput, low queue sizes and small flow completion times. To show this, we present a simple experiment.

\noindent
\paragraph{\textbf{Setup.}}
We compare \gls{tcp} NewReno (loss-based), \gls{dctcp} (ECN-based), and \proto (\service service-based) using a single bottleneck topology (10~Gbps links, 100$\mu s$ (RTT), and 250~packets buffer).
In this setup, five senders generate background traffic and 10 senders send a burst of shortflows ($50KB$) every $0.1~s$.
We evaluate this scenario using NS3~\cite{ns3} simulations. 

\begin{figure}[t]
    \centering
    \vspace{-0.1in}
    \begin{subfigure}[b]{0.9\linewidth}
        \centering
        \begin{tikzpicture} [scale=0.8]
            \begin{axis}[legend style={at={(0.16,.98)},anchor=north,legend cell align=left},
                xlabel=Goodput (Mbps), height=4.cm, width=8cm,
                ylabel=CDF (goodput),y label style={at={(0.06,0.4)}},
                ymajorgrids=true,xmajorgrids=true,]
                \addplot[color=black, densely dotted, line width=2.0pt] table [x=x, y=y, col sep=comma] {figures/motivation/newreno/goodput_cdf.csv};
                \addlegendentry{\scriptsize TCP}
                \addplot[color=blue, loosely dashed, line width=2.0pt] table [x=x, y=y, col sep=comma] {figures/motivation/dctcp/goodput_cdf.csv};
                \addlegendentry{\scriptsize DCTCP}
                \addplot[color=red, line width=2.0pt] table [x=x, y=y, col sep=comma] {figures/motivation/dwtcp/goodput_cdf.csv};
                \addlegendentry{\scriptsize DWTCP}
            \end{axis}
        \end{tikzpicture}
        \caption{Goodput}
        \label{FIG:motivation_goodput}
    \end{subfigure}
    \begin{subfigure}[b]{0.9\linewidth}
        \centering
        \begin{tikzpicture} [scale=0.8]
            \begin{axis}[legend style={at={(0.5,1.1)},legend columns=-1,anchor=north,legend cell align=left},
                xlabel=Time (s), height=4.cm, width=8cm,
                ylabel=Queue length (Packet),y label style={at={(0.06,0.4)}},
                ymajorgrids=true,xmajorgrids=true,]
                \addplot[color=black, densely dotted, line width=2.0pt] table [x=x, y=y, col sep=comma] {figures/motivation/newreno/packetsInQueue-r0-q0.csv};
                \addlegendentry{\scriptsize{TCP}}
                \addplot[color=blue, loosely dashed, line width=2.0pt] table [x=x, y=y, col sep=comma] {figures/motivation/dctcp/packetsInQueue-r0-q0.csv};
                \addlegendentry{\scriptsize{DCTCP}}
                \addplot[color=red, mark=., line width=2.0pt] table [x=x, y=y, col sep=comma] {figures/motivation/dwtcp/packetsInQueue-r0-q0.csv};
                \addlegendentry{\scriptsize{DWTCP}}
            \end{axis}
        \end{tikzpicture}
        \caption{Queue length}
        \label{FIG:motivation_qlen}
    \end{subfigure}
    \begin{subfigure}[b]{0.9\linewidth}
        \centering
        \begin{tikzpicture}[scale=0.8]
            \begin{axis}[
                ybar,
                ymin=0,
                width=8cm,
                height=4cm,
                bar width=8pt,
                ylabel=FCT (ms),y label style={at={(0.1,0.52)}},
                xlabel=Percentile,
                symbolic x coords={Mean, $90^{th}$, $99^{th}$},
                xtick = data,
                enlarge y limits={value=0.2,upper},
                enlarge x limits=0.25,
                ymajorgrids=true,xmajorgrids=true,
                legend style={at={(0.5,.99)},legend columns=-1,anchor=north,legend cell align=left, fill=white, fill opacity=0.6, draw opacity=1, text opacity =1},
            ]
            \addplot[pattern color=black, pattern=crosshatch] coordinates {(Mean, 3.21355176) ($90^{th}$, 5.85965) ($99^{th}$, 7.3801244)};
            \addplot[pattern color=blue, pattern=north east lines] coordinates {(Mean, 2.56561561) ($90^{th}$, 3.301544) ($99^{th}$, 3.77181)};
            \addplot[pattern color=blue, pattern=dots] coordinates {(Mean, 2.29605418) ($90^{th}$, 2.733726) ($99^{th}$, 2.9990939)};
            \addplot[fill=red] coordinates {(Mean, 1.14499961) ($90^{th}$, 1.349249) ($99^{th}$, 1.3770419)};
            \legend{\scriptsize TCP, \scriptsize DCTCP, \scriptsize DCTCP (K=7) , \scriptsize DWTCP}
            \end{axis}
        \end{tikzpicture}
        \caption{Flow completion time}
        \label{FIG:motivation_fct}
        \vspace{-0.1in}
    \end{subfigure}
       \caption{Early congestion feedback helps maintain (a) high goodput, (b) low queue size, and (c) small flow completion times. \proto uses congestion signal from low priority queues and thus adapts to network conditions much quickly and efficiently.}
       \label{FIG:motivation}
       \vspace{-0.2in}
\end{figure}
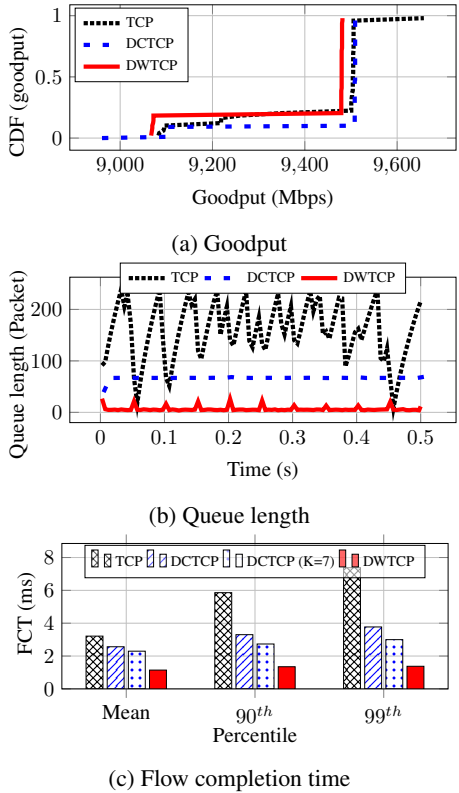

\noindent
\paragraph{\textbf{Signals from High Priority Queue.}}
Loss-based CC schemes rely on the state of the queues and detect congestion when the network is already congested. 
As a result, the senders using TCP NewReno reduce their congestion window aggressively, creating large variations in the queue occupancy. Although, this results in higher throughput for the long flows, it impacts short flow latency as they have to wait longer in the queues as shown in Figure~\ref{FIG:motivation_qlen}.

ECN alleviates some challenges associated with the loss-based schemes by providing early indication of congestion.
However, its performance depends on the queue size. Figures~\ref{FIG:motivation_goodput} and \ref{FIG:motivation_qlen} show that DCTCP can maintain high throughput and low queues (around its threshold) respectively. However, the completion time of short flows depends on the ECN threshold (Figure \ref{FIG:motivation_fct}). 

\noindent
\paragraph{\textbf{Signals from Low Priority Queue.}}
From Figure~\ref{FIG:motivation_qlen}, we can observe that feedback from the low priority queue helps in detecting congestion earlier than both ECN and loss, hence maintaining small queues.
This allows the sender to be aggressive in its control laws (proportional to link busyness).
Note that all the other protocols observe packet drops when a burst of short flows arrives. In contrast, \proto maintains low queue length, therefore, it has enough room to converge without causing extra latency or packet drop. In addition, \proto achieves similar goodput as TCP NewReno and DCTCP, even though it does not rely on queue occupancy for high throughput (Figure~\ref{FIG:motivation_goodput}).

\subsection{Design Space}
A key design question is how to get feedback from the low priority queues and what congestion signals can be inferred.
In this work, we propose using \service packets to observe the state of the low priority queue, in terms of bandwidth availability and link busyness.
\service shares similarities with the probing-based schemes that observe the link state~\cite{anderson2006pcp, eccp, bart, pathChirp, topp}. However, it is fundamentally different from these approaches as it uses low priority queues to probe the network.
Using low priority allows \service to operate without hurting the normal traffic, unlike existing schemes.
\service packets might cause a small delay for normal traffic if the later arrives while serving the former.
This is reduced by using small packet size (64~bytes), and can be reduced further by using packet preemption (IEEE 802.1Qbu).

An orthogonal design choice could be to send actual data packets in the low priority queue as \service. Such an approach would be useful for protocols that open multiple communication channels such as MPTCP~\cite{mptcp}, Stream Control Transmission Protocol (SCTP)~\cite{sctp} and HTTP~1.1~\cite{http1_1}. However, for single channel protocols, it may create issues with the out-of-order delivery and timeouts.
We leave the evaluation of such an approach as future work.
    
Another design choice could be to measure the bandwidth availability and link busyness at the switch and share such information with the end-hosts using a mechanism similar to INT.
However, such a design requires changes to the switch, which is orthogonal to our design objectives. 

\paragraph{\textbf{Why not ECN with Small Threshold?}} A simple design choice could be to use DCTCP with a very small marking threshold. 
However, as shown in the original DCTCP paper, small marking thresholds results in severe underutilization for the long flows~\cite{alizadeh2010data}. 
To further understand this, we repeat the experiment in \S~\ref{sec:benefits} and set marking threshold ($K$) to seven packets (equivalent to the buffer occupancy of \proto). we note that although setting a smaller threshold for DCTCP improves the completion time of the short flows (see Figure\ref{FIG:motivation_fct}), it is still worse than \proto which takes advantage of early and amplified congestion signals and is more aggressive in its control laws. 
%
The long flows also experience under-utilization with DCTCP in this scenario. 
We observe a similar trend in the testbed evaluation (See \S~\ref{sec:testbed}).

\begin{figure}[t]
    \centering
    \begin{tikzpicture} [scale=0.9]
        \def\qlen{1.5}
        \def\qheight{0.25}
        \def\qxshift{0.2}
        \def\qyshift{0.2}
        \node [circle,draw=black, fill=white!50!gray, inner sep=0pt,minimum size=7pt] (qserver) at (0,0) {+};
        \foreach \i in {1,-1}
        {
            \draw[black] (-\qlen-\qxshift, \i*\qyshift) -- (-\qxshift, \i*\qyshift) -- (-\qxshift, \i*\qyshift+\i*\qheight) -- (-\qlen-\qxshift, \i*\qyshift+\i*\qheight) node[pos=1, yshift=5] {}; 
        }
        \node[yshift=5] at (-0.5*\qlen-\qxshift, \qyshift+\qheight) {HPQ}; 
        \node[yshift=-5] at (-0.5*\qlen-\qxshift, -\qyshift-\qheight) {LPQ}; 
        
        \draw[black, ->] (-\qxshift, \qyshift+0.5*\qheight) -- (qserver); 
        \draw[black, ->] (-\qxshift, -\qyshift-0.5*\qheight) -- (qserver); 
        \draw[black, ->] (qserver) --  (2*\qxshift, 0); 
        \def\pktheight{0.25}
        \def\pktwidth{0.15}
        \def\pktspace{0.05}
        \def\pktxshift{-1.2}
        \def\pktyshift{0.15}
        \foreach \i in {1,2,3,4,5}
        {
            \draw[pattern color=black, pattern=crosshatch] (\i*\pktspace+\i*\pktwidth-\pktwidth+\pktxshift,\pktspace+\pktyshift) rectangle (\i*\pktspace+\i*\pktwidth+\pktxshift,\pktspace+\pktheight+\pktyshift) node[pos=.5] (p\i) {};
        }
        \def\pktxshift{-0.8}
        \def\pktyshift{-0.5}
        \foreach \i in {1,2,3}
        {
            \draw[pattern color=black, pattern=north east lines] (\i*\pktspace+\i*\pktwidth-\pktwidth+\pktxshift,\pktspace+\pktyshift) rectangle (\i*\pktspace+\i*\pktwidth+\pktxshift,\pktspace+\pktheight+\pktyshift) node[pos=.5] (p\i) {};
        }
        \def\xshift{-1}
        \def\switchdim{2}

        \draw[black] (-.5*\switchdim+\xshift,-0.5*\switchdim) rectangle (0.5*\switchdim+0.5\xshift,0.5*\switchdim) node[pos=.5,yshift=40pt] (sw) {};
        \node at (sw) [scale=0.15, xshift=5cm,yshift=-4.5cm] {\router{}};
        \def\nodewidth{2}
        \def\nodespacing{2}
        \def\nodeheight{2}
        \draw[black] (-\nodespacing-\nodewidth+\xshift,0.5*\switchdim) rectangle (-\nodespacing+\xshift,\nodeheight+0.5*\switchdim) node[pos=.5,yshift=1.3cm] (src) {Source};
        \draw[black] (\nodespacing+\xshift,+0.5*\switchdim) rectangle (\nodespacing+\nodewidth+\xshift,\nodeheight+0.5*\switchdim) node[pos=.5,yshift=1.3cm] (dst) {Destination};

        \def\boxheight{0.6}
        \def\boxspacing{0.1}
        \def\srcxleft{-\nodespacing-\nodewidth+\xshift+\boxspacing}
        \def\srcxright{-\nodespacing+\xshift-\boxspacing}
        \def\hosty{\nodeheight+0.5*\switchdim}
        \draw[black] (\srcxleft,\hosty-0.1*\boxheight) rectangle (\srcxright+0.5*\boxspacing,\hosty-1*\boxheight) node[pos=.5,] (srcapp) {Application};
        \draw[black] (\srcxleft+\boxspacing+0.5*\nodewidth,\hosty-1*\boxheight-\boxspacing) rectangle (\srcxright,\hosty-2.4*\boxheight-\boxspacing) node[pos=.5,] (tcp) {TCP};
        \draw[black] (\srcxleft, \hosty-1*\boxheight-\boxspacing) -- (\srcxleft+0.5*\nodewidth, \hosty-1*\boxheight-\boxspacing) -- (\srcxleft+0.5*\nodewidth, \hosty-2.5*\boxheight-\boxspacing) -- (\srcxleft+0.25*\nodewidth,\hosty-2.5*\boxheight-\boxspacing) -- (\srcxleft+0.25*\nodewidth,\hosty-1.5*\boxheight-\boxspacing) -- (\srcxleft,\hosty-1.5*\boxheight-\boxspacing) -- (\srcxleft, \hosty-1*\boxheight-\boxspacing) node[pos=0.5, xshift=0.5cm] (sa) {\scriptsize Analyzer};
        \draw[black] (\srcxleft,\hosty-1.5*\boxheight-1.3*\boxspacing) rectangle (\srcxright-0.67*\nodewidth,\hosty-2.5*\boxheight-1*\boxspacing) node[pos=.5, xshift=5pt,text width=0.8cm] (sg) {\baselineskip=3pt\tiny Scout Gen\par};

        \draw[black] (\nodespacing+\xshift+\boxspacing,\hosty-0.1*\boxheight) rectangle (\nodespacing+\nodewidth+\xshift-\boxspacing,\hosty-1*\boxheight) node[pos=.5,] (dstapp) {Application};
        \draw[black] (\nodespacing+\xshift+\boxspacing,\hosty-1*\boxheight-\boxspacing) rectangle (\nodespacing+\nodewidth+\xshift-\boxspacing,\hosty-2*\boxheight-\boxspacing) node[pos=.5] (sr) {\scriptsize Scout Reflector};

        \draw[line width=0.5mm] (\srcxleft+0.5*\nodewidth, 0.5*\switchdim) -- (\srcxleft+0.5*\nodewidth, 0) -- (-1*\switchdim, 0);
        \draw[line width=0.5mm] (\nodespacing+\nodewidth+\xshift-0.5*\nodewidth, 0.5*\switchdim) -- (\nodespacing+\nodewidth+\xshift-0.5*\nodewidth, 0) -- (0.25*\switchdim, 0);

        \draw[thick, blue, ->] (tcp) to[out=270,in=180] ($ (qserver) + (0,1.5*\qyshift) $) to[out=0,in=270] ($ (sr) + (-0.2,-0.5*\boxheight) $);
        \draw[thick, red, ->] (sg) to[out=270,in=180] ($ (qserver) + (0,-1.5*\qyshift) $) to[out=0,in=270] ($ (sr) + (0.2,-0.5*\boxheight) $);
        \draw[thick, dashed, red, ->] (sr) to[out=270,in=0] (qserver) to[out=180,in=270] ($ (sa) + (0.2,-0.2) $);
    \end{tikzpicture}
    \caption{\service service overview}
    \label{FIG:ss}
    \vspace{-0.2in}
\end{figure}
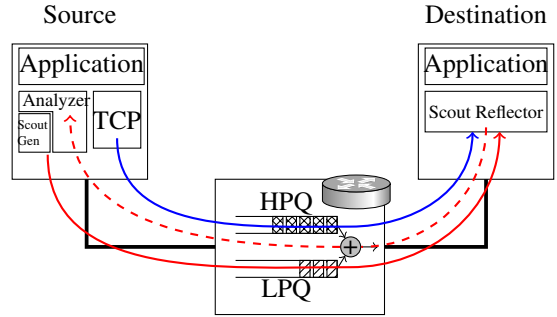

\vspace{-0.1in}
\section{\service}
\label{sec:scout}

\subsection{Overview}
\service is a probing-based congestion signaling mechanism that has three main components as depicted in Figure~\ref{FIG:ss}.

\noindent
\paragraph{\textbf{\service Generator.}}
A \service sender sends very small (64~bytes) low priority \service packets alongside the normal TCP traffic to observe the network state. 
%
%
\service packets can be sent using either a parallel UDP socket, or on the same TCP socket after setting priority as a flag to distinguish \service traffic. 
Using UDP socket to send \service requires updating hashing at the switches to make sure the \service traffic takes the same path as the data packets. However, using the same TCP socket allows \service packets to be served through the same datapath in case of using ECMP. 

\noindent
\paragraph{\textbf{\service Reflector.}}
\service reflector acts as a server that echoes back the ACKs for \service packets.
It can be implemented using IPTABLEs without requiring any changes in the TCP stack.

\noindent
\paragraph{\textbf{\service Analyzer.}}
\service analyzer is the brain of \service that has two main responsibilities. 
First, it analyzes the \service packets and the \service ACKs to extract congestion signals. 
Second, it interacts with the congestion control module to react based on the observed network state.

\service service does not require any changes in the switches. It can be implemented in commodity switches using a strict priority queue. 
At the switch side, normal traffic is served with strict high priority, while \service traffic is served with the low priority.
Because \service packets are sent with low priority, switches drop them in case of contention, which guarantees that the \service incurs very little overhead in the network (less than 0.1\% in our experiments).
%
Moreover, \service does not require switches to reserve too much buffer capacity for low priority queue. Even using 600~bytes (5-10, 64-byte \service packets) for the low priority queue provides the best performance results.

\subsection{\service for Congestion Signaling}
\service provides three key benefits for the congestion control design.
First, it helps in detecting available bandwidth in the network.
Second, it helps in determining the amount of congestion (in the form of packet loss and delay).
Third, it provides this information earlier than existing mechanisms.

\noindent
\paragraph{\textbf{Bandwidth Availability.}}
\service traffic, at the switches, can only fill in the gaps between normal traffic packets (See Figure~\ref{FIG:scoutAffect}). 
Once an ACK for the \service packet of size $S$ is received, it provides an implicit guarantee from the network that all switches through the datapath can allocate at least $S$ bytes for that flow. We call such a signal “Bandwidth Availability”.
%
%
The transport protocol can increase its window size faster, \emph{proportional} to $S$, to utilize any available bandwidth. 
Note that, \service is not trying to estimate the exact value of the available bandwidth, rather it tries to get implicit permission from the network to use a certain limit of the extra bandwidth through the datapath.

\paragraph{\textbf{Link Busyness.}}
Link busyness indicates the amount of time that the switch is continously busy serving normal traffic. Increase in link busyness, {\em i.e.} smaller gaps between normal packets, is an early indication of congestion. \service enables an end-host to detect link busyness using the delay and loss of \service packets in the network. 
%
%
For example, delay ($d_s$) in a \service packet (and corresponding ACK) can be used as an early indication of congestion.
Similarly, loss of a \service packet (or ACK) indicates severe congestion and the potential of a packet loss in the network.
The sender can use these signals to reduce congestion window and prevent queue buildups and packet drops in the network. 

\begin{figure}[t]
    \centering
    \begin{tikzpicture} [scale=0.8], font=\large]
        \begin{axis}[legend style={at={(0.3,.8)},anchor=north,legend cell align=left},
            xlabel=Link Utilization (\%), height=4.cm, width=8cm,
            ymajorgrids=true,xmajorgrids=true,
            ymin=0.2,ymax=1.1]
            \addplot[color=red, mark=o, line width=2.0pt, each nth point=5] table [x=util, y=d_s, col sep=comma] {figures/correlations/rate_ds_inv.csv};
            \addlegendentry{$d_s$ (ms)}
            \addplot[color=black, mark=square, line width=2.0pt, each nth point=5] table [x=util, y=d_inv, col sep=comma] {figures/correlations/rate_ds_inv.csv};
            \addlegendentry{$\tau / d_s$ (Ratio)}
        \end{axis}
    \end{tikzpicture}

    \caption{Relation between the \service delay and link utilization. Ratio of RTT to \service delay ($\tau / d_s$) provides a very clear indication of the congestion in the network. }
\label{FIG:dynamics:scoutRTT}
\vspace{-0.2in}
\end{figure}
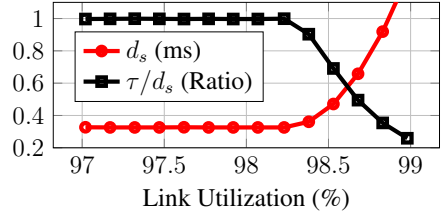

\noindent
\paragraph{\textbf{Early Congestion Detection and Signal Amplification.}}
\service can detect link busyness several Round-Trip times (RTTs) (denoted by $\tau$ here) before the link reaches full utilization, which is very fundamental in designing congestion control algorithms.
%
Figure~\ref{FIG:amplification_linkvsdelay} shows that the delay ($d_s$) encountered by \service packets gets several times higher than the delay of normal traffic (5-15x) as the load increases.
Furthermore, Figure~\ref{FIG:dynamics:scoutRTT} depicts the relation between \service delay ($d_s$) and link utilization. For a fixed \service packet size, \service throughput is inversely proportional to \service delay.  As depicted with the black line, the normalized inverse \service delay ($\tau/d_s$) provides a clear indication of congestion as the link utilization approaches 100\%.
Similarly, Figure~\ref{FIG:amplification_timeToFullQueue2} shows the time difference in how quickly \service can detect the packet drop based on the aggressiveness (increase in rate). A \service loss predicts link is full several (tens to hundreds of) RTTs earlier than similar signals in current loss-based or delay-based schemes.

\subsection{\service Optimizations}
%
In this work, we propose the following optimizations to improve congestion signaling timeliness and accuracy, to minimize overhead and switch buffer requirements, and to increase scalability.

\noindent
\paragraph{\textbf{\service Packets Injection Rate ($x$).}}\label{SEC:scoutRate}
The rate at which \service packets are injected directly affects the timescale at which congestion can be detected in the network and the overhead due to \service traffic.
Too low \service rate (one packet every few RTTs) can minimize the overhead, but it can also severely impact the timeliness of getting congestion state of the network, and ability to estimate the available bandwidth. 
Too high \service rate (on sub-RTT level) can cause significant overhead, while not providing significant benefits for congestion detection. 
%
We propose to send one \service packet, per flow, every base RTT of the network as a baseline. This also helps with requiring smaller buffers to hold \service packets in the low priority queues. We show the impact of different injection rates in the evaluation.

\noindent
\paragraph{\textbf{Dynamic \service Coefficient for Bandwidth Availability Estimate.}}
Sending one \service per RTT limits the ability to detect available bandwidth in the network. 
Instead of using large \service packet size, or varying the \service injection rate, we use a dynamic \emph{\service coefficient} $\alpha$. We send a small \service packet (of 64 bytes) every RTT, yet dynamically scale $\alpha$ to adjust bandwidth estimate associated with an incoming \service packet. 
We double the value of $\alpha$ upon receiving a \service ACK; and reduce the $\alpha$ by half upon observing a \service packet loss.
The congestion window of the flow is increased according to the \service coefficient, instead of the \service packet size, upon receiving a \service ACK.
This provides significant improvement in convergence rate and helps adapt to network conditions in case of congestion.

\noindent
\paragraph{\textbf{No \service for Flows Smaller than Initial Congestion Window Size.}}
We do not use \service for a flow if its size is smaller than the initial congestion window size. The intuition here is that if a flow is not going to last for more than one RTT, then getting network state feedback for such a flow is not useful. Despite not using \service for such short flows, we can still see improvements in highly skewed workloads like datamining\cite{greenberg2009vl2} since \service can keep queue occupancies low and therefore short flows can complete faster. 
    Moreover, it significantly, reduces the CPU and network overheads, due to \service packets, for such workloads.

\noindent
\paragraph{\textbf{Per-datapath \service.}}
To further reduce \service traffic and increase scalability, \service packets can be sent between each pair of hosts (per-datapath), instead of per-flow, and the received congestion signal can be distributed among all active flows between each pair. For example, bandwidth availability signal can be distributed equally among all active flows, and link busyness can be conveyed to all active flows.
Our testbed evaluation shows that this strategy does not affect the performance, yet reduces the \service overhead significantly.
%
We note that, our main design and evaluation does not assume this optimization.

\subsection{\service Overheads Comparison}

\service packets do not incur significant overheads in terms of CPU utilization at the end-host and/or extra traffic (\service) in the network. To understand the overheads, let us assume that we are sending $x$ \service packets per RTT for each connection. 

For the CPU overhead, this setting will require generating $x100k$ \service packets per second per flow for a datacenter with $10\mu$s RTT (as in a 100G datacenter network). If the end-host has 100 flows, then it will need to generate $\approx$ $x$10 million \service per second, which can be generated using one core in commodity servers. 
Similarly, as for the network overhead, for each \service packet of 64 bytes, the host also sends a number of packet equal to its \gls{cwnd}.
For a typical MTU setting of 1500 bytes, this means $CWND * 1500$ bytes. Hence, the worst-case scenario can be calculated when $CWND=1$ to be $64/1500=4.2\%$.

In practice, however, we do not observe this extreme case overhead. For example, Figure~\ref{FIG:motivation_goodput} shows that the \proto long flow has less than $0.2\%$ overhead in terms of goodput.
This is for the following reasons:
First, the \service packets are served only when there is available bandwidth in the network. The LPQ \service packets lose to the HPQ dat packets when the link is highly utilized.
Second, the average $CWND$ is much larger than $1$. 
Lastly,
the per-path implementation of the \service scales down the overhead by the multiplexing factor per connection (\emph{e.g.}, in the worst case, for a load of 100 flows per server, and an average multiplexing of 10 flows per path, this leads to at most 0.5\% overhead due to the \service packets).

On the other hand, the overhead of existing schemes such as INT for a usual 5-hop topology is reported to be around 6.8\%~\cite{ben2020pint}. The most recent probabilistic version of the INT telemetry method~\cite{ben2020pint} 
tries to minimize the per packet overhead by probabilistically distributing the congestion signal over the packets of a single flow.
This potentially reduces the per packet overhead to be as small as 2 bytes which makes the overall overhead smaller than 0.2\%.
However, it should be noted that due to the distributed nature of the signal over the packets, only large flows can benefit from 
the PINT signal and shorter flows still encounter the normal INT overhead~\cite{ben2020pint}. 

\section{\proto}
\label{dwtcp}
\subsection{Overview}
\gls{dwtcp} uses the \service bandwidth availability, \service delay and \service loss signals to control the congestion window. 
At its base, \proto uses TCP as a transmission control mechanism.

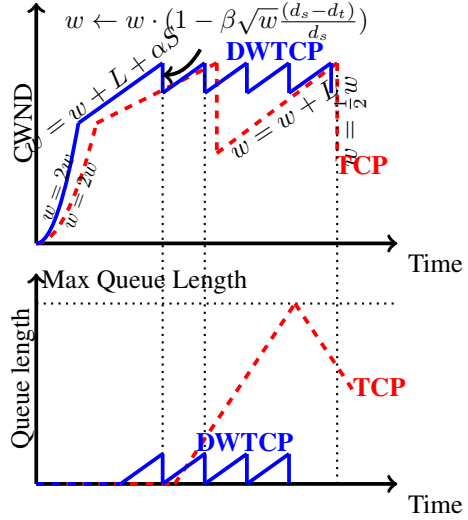
\begin{figure}[t]
    \centering
    \begin{tikzpicture} [scale=0.8]
        \def\xmax{6}
        \def\ymax{4}
        \draw[line width=0.5mm,->] (0,0) -- (\xmax,0) node[anchor=north west] {Time};
        \draw[line width=0.5mm,->] (0,0) -- (0,\ymax) node[midway, rotate=90, yshift=5pt] {CWND};
        \draw[line width=0.5mm,red, dashed] (0,0) parabola (1,2) node[black, midway, rotate=70,xshift=1cm, yshift=-0.4cm] {$w=2w$};
        \draw[line width=0.5mm,red, dashed] (1,2) -- (3,3) node[black, midway, rotate=25, xshift=-10pt, yshift=-5pt] {}; 
        \draw[line width=0.5mm,red, dashed] (3,3) -- (3,1.5) node[black, midway, rotate=90, xshift=-10pt, yshift=7pt] {};
        \draw[line width=0.5mm,red, dashed] (3,1.5) -- (5,3) node[black, midway, rotate=33, yshift=-5pt] {$w=w+L$};
        \draw[line width=0.5mm,red, dashed] (5,3) -- (5,1.5) node[black, midway, rotate=90, xshift=-5pt, yshift=-5pt] {$w=\frac{1}{2}w$};
        \node[red] at (5.4, 1.3) {\textbf{TCP}};
        \def\xshort{0.3}
        \def\xstep{0.7}
        \def\ystep{0.5}
        \draw[line width=0.5mm,blue] (0,0) parabola (1-\xshort,2) node[black, midway, rotate=70,xshift=1cm, yshift=0.cm] {$w=2w$};
        \draw[line width=0.5mm,blue] (1-\xshort,2) -- (3-3*\xshort,3) node[black, midway, rotate=37, xshift=-5pt, yshift=5pt] {$w=w+L+\alpha S$};
        \draw[line width=0.5mm,blue] (3-3*\xshort,3) -- (3-3*\xshort,3-\ystep);

        \foreach \i in {0, 1,2,3}
        {
            \draw[line width=0.5mm,blue] (3-3*\xshort+\xstep+\i*\xstep,3) -- (3-3*\xshort+\xstep+\i*\xstep,3-\ystep);
            \draw[line width=0.5mm,blue] (3-3*\xshort+\i*\xstep,3-\ystep) -- (3-3*\xshort+\xstep+\i*\xstep,3);
        }

        \node[blue] at (4, 3.2) {\textbf{\proto}};
        \node(eq) at (3, 3.7) {$w \gets w \cdot (1 - \beta \sqrt{w} \frac{(d_s - d_{t})}{d_s})$};
        \draw[line width=0.5mm, ->] (eq) to[out=240,in=0] (2.1, 2.8);

        \draw[line width=0.5mm,->] (0,-\ymax) -- (\xmax,-\ymax) node[anchor=north west] {Time};
        \draw[line width=0.5mm,->] (0,-\ymax) -- (0,-.5) node[midway, rotate=90, yshift=5pt] {Queue length};
        \draw[thick, dotted] (2.1, 2.8) -- (2.1, -\ymax);
        \draw[thick, dotted] (2.1+\xstep, 2.8) -- (2.1+\xstep, -\ymax);
        \draw[thick, dotted] (5, 2.8) -- (5, -\ymax);
        \draw[thick, dotted] (0, 3 -\ymax) -- (\xmax,3 -\ymax) node[pos=0.3, above ] {Max Queue Length};

        \def\xstart{2.3}
        \draw[line width=0.5mm,red, dashed] (0.5,-\ymax) -- (\xstart,-\ymax);
        \draw[line width=0.5mm,red, dashed] (\xstart,-\ymax) -- (\xstart+2,3-\ymax);
        \draw[line width=0.5mm,red, dashed] (\xstart+2,3-\ymax) -- (\xstart+3.,1.5-\ymax) node[red, anchor=north west, xshift=-5pt, yshift=10pt] {\textbf{TCP}};

        \draw[line width=0.5mm,blue, dashed] (0,-\ymax) -- (1.7-\xshort,-\ymax);
        \foreach \i in {0, 1,2,3}
        {
            \draw[line width=0.5mm,blue] (1.7-\xshort+\i*\xstep,-\ymax) -- (1.7-\xshort+\xstep+\i*\xstep,\ystep-\ymax);
            \draw[line width=0.5mm,blue] (1.7-\xshort+\xstep+\i*\xstep,\ystep-\ymax) -- (1.7-\xshort+\xstep+\i*\xstep,-\ymax);
        }
        \node[blue] at (3.5, \ystep-\ymax + 0.2) {\textbf{\proto}};
    \end{tikzpicture}
    \caption{\proto work mechanism. Early detection and early reaction to congestion make \proto stable, converge faster, and maintain smaller queues.}
    \label{FIG:workMechanism}
    \vspace{-0.1in}
\end{figure}

At high level, the interaction between \service and \gls{tcp} algorithm is as follows:
\begin{itemize}
    \setlength{\itemsep}{0pt plus 1pt}
    \item When a \gls{tcp} connection is established, the \service connection also starts sending small \service packets periodically from the sender side towards the receiver and waits for their ACKs.
    \item Scout packet ACKs, from the \service receiver, work as an indication of \textit{bandwidth availability}. \proto increases \gls{cwnd} by a factor $\alpha$ of the number of received Scouts: $w \leftarrow w + \alpha \cdot S$, as shown in Figure~\ref{FIG:workMechanism}, where $S$ is the \service packet size and $\alpha$ is the \service window size as described in \S\ref{SEC:scoutRate}.
    \item In addition, \proto  also increases the congestion window by $L$ upon receiving a data packet ACK, similar to TCP.
    \item \service delay $d_s$ is used as an indication of link busyness. If $d_s$ is larger than a target delay $d_t$ (as the maximum delay after which a \service packet is considered lost in the network; more on this coming up), \proto connection makes a non-aggressive rate reduction decision. Such an early signal allows \proto to react quickly to congestion and maintain lower queue length as shown in Figure~\ref{FIG:workMechanism}.
\end{itemize}

The basic signal in \proto is the \service delay, which helps the source decide when to increase or decrease congestion window.
Unlike delay-based congestion control algorithms~\cite{swift2020}, \service does not need special hardware at the host to measure accurate time stamps for $d_s$ calculation. 

\subsection{\proto Control Algorithm}
\label{SEC:dynamics}

\noindent
\paragraph{\textbf{\proto Control Laws.}}
\proto uses the following control laws to manage the congestion window (CWND) of its flows.

\begin{equation}
    w \gets
    \begin{cases}
            w + L + \alpha S & if d_s \le d_{t} \\
            w \cdot (1 - \beta \sqrt{w} \frac{(d_s - d_{t})}{d_s}) & Otherwise
    \end{cases}
    \label{EQ:dynamics}
\end{equation}

Here, $d_s$ is the \service delay, $d_t$ is the target \service delay, $w$ is the current congestion window, $L$ is the data packet size, $S$ is the \service packet size, $\beta$ is a constant used to limit the maximum window reduction to $1/2$ (similar to TCP) and $\alpha$ is the dynamic \service coefficient that controls the aggressiveness of \proto (as described in \S\ref{SEC:scoutRate}). In addition, similar to \cite{swift2020}, we multiply the decreasing signal by the square root of the current congestion window. Such an approach enhances fairness between flows as it forces the flows with larger $w$ to reduce their congestion window faster.

\noindent
\paragraph{\textbf{\proto Control Paramenters.}}
\proto performance is controlled by 
the target \service delay $d_t$, \service coefficient $\alpha$, and the constant $\beta$.

First, $d_t$ along with the $d_s$ helps decide when to switch between the increasing phase and the decreasing phase.
Second, their ratio $\big( \frac{d_s - d_{t}}{d_s} \big)$ is used to dynamically decrease the congestion window proportional to the network state. Higher the delay, higher the decrement value is.

In \proto, we define \service delay $d_s$ as follows:

\begin{equation*}
    \begin{split}
        d_s = max(&scout.sendTime - scout.rcvTime, \\
        & Now - unackedScout.sendTime)
    \end{split}
\end{equation*}

where $scoutSendTime$ and $scoutRcvTime$ are the transmission and receiving timestamps of the last \service respectively.
The intuition here is to use the \service throughput as an indication of \textit{link busyness}.
In other words, if the queue is busy serving normal traffic and the \service packets are congested, it is a clear indication that flows must reduce their congestion window.

\proto uses target delay $d_{t}$ as the maximum delay after which a \service packet is considered lost in the network. 
$d_t$ depends on the network RTT ($\tau$) and the link capacity.
\proto sets $d_t$ as: 

\begin{equation}
    d_{t} \gets \tau + K \cdot \frac{L}{C},
\label{EQ:dtarget}
\end{equation}

where $L$ is the packet length, $C$ is the line rate, and $K$ indicates how long the high priority queue was busy before allowing \service to be served in the low priority queue in data packet time unit. In our experiments, we set $K$ to 100. The intuition here is that by allowing the switch to serve one \service packet every $K$ data packets, we limit the link utilization to $K/(K+S) \ge 99\%$. 

We suggest using $\beta \approx \frac{1}{2 \sqrt{BDP / N}}$, where $BDP$ is the bandwidth-delay product of the network and $N$ is an estimation of the average flow number. However, such an estimation does not need to be accurate as shown by the stability condition deduced in Appendix \ref{SEC:analysis}.\\

\noindent
\paragraph{\textbf{\proto Stability Analysis.}}
We  mathematically analyze the stability of \proto and demonstrate that (proof in Appendix \ref{SEC:analysis}):

\begin{theorem}
    \proto is spiral stable and has one stable point (at ${Q=0, w=BDP/N}$) as long as the number of processed packet in the high priority queue before serving any \service $(\bar{k})$ is less than $2 \frac{BDP}{\sqrt{\beta}}$.
\end{theorem}

\begin{algorithm}[t]
    \DontPrintSemicolon
    \SetKwFunction{FMain}{AckRcvd}
    \SetKwProg{Fn}{Function}{:}{}
    \KwData{Data ACK, Scout ACK, Scout delay ($d_s$), $\alpha$, $\beta$ }
    \KwResult{New Congestion window $w$}
    $w \gets w_{init}$\;
    $T =$ physical hardware RTT\;
    $d_{t} \gets \tau + K \cdot \frac{L}{C}$\;
    $DET = Now()$ \tcp*{Decrease Execution Time}
    $\beta = 1/16$\;
    $\alpha = 1$\;
    \hrulefill\;
    \Fn{\FMain{$w$, $ACK.TYPE$}}{
        \tcc{Increasing phase}
        $l = (ACK.TYPE == Data)?\ L\ :\ 0$\;
        $s = (ACK.TYPE == Scout)?\ S\ :\ 0$\;
        \uElseIf {TCP.state == Slow start}{
            $w \gets 2 \cdot w + \alpha \cdot s$;
        }
        \Else{
            $w \gets w + l + \alpha \cdot s$;
        }
        return $w$\;
    }
    \hrulefill\;
    \SetKwFunction{FScoutDelayed}{ScoutDelayed}
    \Fn{\FScoutDelayed{$w$, $scout.sendTime$, $scout.rcvTime$, $unackedScout.sendTime$}}{
        $d_s = max(scout.sendTime - scout.rcvTime, Now - unackedScout.sendTime)$\;
        \uIf{$ds > d_{t}$}{
            \uIf {$Now() >= DET + T $}{
                \tcc{Decreasing phase}
                $TCP.ssthresh = w$\;
                $w \gets w \cdot \max{\big((1 - \beta \sqrt{w} \frac{(d_s - d_{t})}{d_s}), 0.5\big)}$ \;
                $DET = Now()$\;
            }
        }
        return $w$\;
    }
    \caption{CWND management using \service}
    \label{ALGO:cwndManagement}
\end{algorithm}

\noindent
\paragraph{\textbf{\proto Algorithm.}}
The working mechanism of \proto is depicted in Algorithm~\ref{ALGO:cwndManagement}. We start by initializing some common variables in lines 1 - 5. The increasing phase, defined by $AckRcvd$ function, is triggered when data or \service acknowledgments are received (Lines 8 - 15). In this function, $l$ and $s$ are initialized to $L$ and $S$ respectively depending on the acknowledge type (Lines 9 and 10). In addition, the \gls{cwnd} is to be increased according to \gls{tcp} state as indicated by lines 12 and 14 for the slow start and congestion avoidance states, respectively. The decreasing phase, defined by $ScoutDelayed$ function, is triggered by not receiving \service for time larger than $d_t$. Thereafter, we calculate the new congestion window based on Equation (\ref{EQ:dw}) as shown in Line 22. We also define a new variable called $DET$ to limit the congestion window reduction to once per RTT (see condition in line 20).

\noindent
\paragraph{\textbf{Using \service Inter-packet Time to Signal Link Busyness.}} Similar to \service delay, we can use \service inter-packet gap ($ipg_{s}$) to enhance the stability and fairness of the system. Hence, we can modify Equation~\ref{EQ:dynamics} as follows:

\begin{equation*}
    w \gets
    \begin{cases}
        w \cdot (1 - 2 \beta \sqrt{w} \frac{(d_s - d_{t})(ipg_{s} - ipg_{t})}{d_s \cdot ipg_{s}}) &if \big((d_s > d_{t})\ \&\\ &(ipg_{s} > ipg_{t})\big) \\
        w \cdot (1 - 2 \beta \sqrt{w} \frac{(ipg_{s} - ipg_{t})^2}{ipg_{s}^2}) & if (ipg_{s} > ipg_{t}) \\
        w \cdot (1 - 2 \beta \sqrt{w} \frac{(d_s - d_{t})^2}{d_s^2}) & if (d_s > d_{t}) \\
        w + L + \alpha S & Otherwise
    \end{cases}
\end{equation*}

where $ipg_{t}$ is the target inter-packet gap which is calculated by dividing $d_{t}$ by \service window size. Because congestion window reduction is limited by $1/2$, we need to accommodate for the square of such a metric by multiplying by $2$ to limit the new metric withing the same range. Our simulation and testbed tests lead to similar results.

\begin{figure*}[htbp]
    \centering
    \begin{subfigure}[b]{0.155\linewidth}
        \begin{tikzpicture}[scale=0.5,font=\LARGE]
            \begin{axis}[legend columns=-1, legend style={at={(1.2,1.18)},anchor=north,legend cell align=left, fill opacity=0.6, text opacity=1}, samples=15,
                xlabel=Time (s), ylabel=Packets, y label style={at={(0.00,0.5)}}, height=5.cm, width=6cm, samples=1,
                ymajorgrids=true,xmajorgrids=true,]
                
                \addplot[color=black, densely dotted, mark=square, line width=2.0pt,each nth point=5] table [x=x, y=y, col sep=comma] {figures/flowsInSequence/newreno/packetsInQueue-r0-q0.csv};
                \addlegendentry{TCP}
                \addplot[color=blue, mark=x, line width=2.0pt,each nth point=5] table [x=x, y=y, col sep=comma] {figures/flowsInSequence/dctcp/packetsInQueue-r0-q0.csv};
                \addlegendentry{DCTCP}
                \addplot[color=red, loosely dashed, mark=x, line width=2.0pt,each nth point=5] table [x=x, y=y, col sep=comma] {figures/flowsInSequence/dwtcp/packetsInQueue-r0-q0.csv};
                \addlegendentry{DWTCP}
            \end{axis}
        \end{tikzpicture}
        \caption{Queue Length}
        \label{FIG:poql}
    \end{subfigure}
    \hspace*{\fill}
    \begin{subfigure}[b]{0.155\linewidth}
        \begin{tikzpicture}[scale=0.5,font=\LARGE]
            \begin{axis}[legend style={at={(0.2,.9)},anchor=north,legend cell align=left, fill opacity=0.6, text opacity=1},
                xlabel=Time (s), ylabel=Jains Index, y label style={at={(0.00,0.5)}}, height=5.cm, width=6cm,
                ymajorgrids=true,xmajorgrids=true,]
                \addplot[color=black, densely dotted, mark=square, line width=2.0pt,each nth point=5] table [x=x, y=y, col sep=comma] {figures/flowsInSequence/newreno/fairness-index.csv};
                \addplot[color=blue, mark=x, line width=2.0pt,each nth point=5] table [x=x, y=y, col sep=comma] {figures/flowsInSequence/dctcp/fairness-index.csv};
                \addplot[color=red, loosely dashed, mark=x, line width=2.0pt,each nth point=5] table [x=x, y=y, col sep=comma] {figures/flowsInSequence/dwtcp/fairness-index.csv};
            \end{axis}
        \end{tikzpicture}
        \caption{Fairness}
        \label{FIG:pof}
    \end{subfigure}
    \hspace*{\fill}
    \begin{subfigure}[b]{0.155\linewidth}
        \centering
        \begin{tikzpicture} [scale=0.5,font=\LARGE]
            \begin{axis}[legend style={at={(0.2,.95)},anchor=north,legend cell align=left, fill opacity=0.6, text opacity=1},
                xlabel=Time (s),ylabel=\huge Rate (Mbps), y label style={at={(0.00,0.5)}}, height=5.cm, width=6cm,
                ymajorgrids=true,xmajorgrids=true,]
                \addplot[color=red, mark=x, line width=2.0pt, each nth point=3] table [x=x, y=y, col sep=comma] {figures/flowsInSequence/newreno/goodput_mbps-0.csv};
                \addplot[color=orange, mark=x, line width=2.0pt, each nth point=3] table [x=x, y=y, col sep=comma] {figures/flowsInSequence/newreno/goodput_mbps-1.csv};
                \addplot[color=blue, mark=x, line width=2.0pt, each nth point=3] table [x=x, y=y, col sep=comma] {figures/flowsInSequence/newreno/goodput_mbps-2.csv};
                \addplot[color=magenta, mark=x, line width=2.0pt, each nth point=3] table [x=x, y=y, col sep=comma] {figures/flowsInSequence/newreno/goodput_mbps-3.csv};
                \addplot[color=black, mark=x, line width=2.0pt, each nth point=3] table [x=x, y=y, col sep=comma] {figures/flowsInSequence/newreno/goodput_mbps-4.csv};
            \end{axis}
        \end{tikzpicture}
        \caption{TCP}
        \label{FIG:performance_overview_TCP}
    \end{subfigure}
    \hspace*{\fill}
        \begin{subfigure}[b]{0.155\linewidth}
        \centering
        \begin{tikzpicture} [scale=0.5,font=\LARGE]
            \begin{axis}[legend style={at={(0.2,.95)},anchor=north,legend cell align=left, fill opacity=0.6, text opacity=1},
                xlabel=Time (s), height=5.cm, width=6cm,
                ymajorgrids=true,xmajorgrids=true,]
                \addplot[color=red, mark=x, line width=2.0pt, each nth point=3] table [x=x, y=y, col sep=comma] {figures/flowsInSequence/dctcp/goodput_mbps-0.csv};
                \addplot[color=orange, mark=x, line width=2.0pt, each nth point=3] table [x=x, y=y, col sep=comma] {figures/flowsInSequence/dctcp/goodput_mbps-1.csv};
                \addplot[color=blue, mark=x, line width=2.0pt, each nth point=3] table [x=x, y=y, col sep=comma] {figures/flowsInSequence/dctcp/goodput_mbps-2.csv};
                \addplot[color=magenta, mark=x, line width=2.0pt, each nth point=3] table [x=x, y=y, col sep=comma] {figures/flowsInSequence/dctcp/goodput_mbps-3.csv};
                \addplot[color=black, mark=x, line width=2.0pt, each nth point=3] table [x=x, y=y, col sep=comma] {figures/flowsInSequence/dctcp/goodput_mbps-4.csv};
            \end{axis}
        \end{tikzpicture}
        \caption{DCTCP}
        \label{FIG:performance_overview_DCTCP}
    \end{subfigure}
\hspace*{\fill}
    \begin{subfigure}[b]{0.155\linewidth}
        \centering
        \begin{tikzpicture} [scale=0.5, font=\LARGE]
            \begin{axis}[legend style={at={(0.2,.95)},anchor=north,legend cell align=left, fill opacity=0.6, text opacity=1},
                xlabel=Time (s), height=5.cm, width=6cm,
                ymajorgrids=true,xmajorgrids=true,]
                \addplot[color=red, mark=x, line width=2.0pt, each nth point=3] table [x=x, y=y, col sep=comma] {figures/flowsInSequence/dwtcp/goodput_mbps-0.csv};
                \addplot[color=orange, mark=x, line width=2.0pt, each nth point=3] table [x=x, y=y, col sep=comma] {figures/flowsInSequence/dwtcp/goodput_mbps-1.csv};
                \addplot[color=blue, mark=x, line width=2.0pt, each nth point=3] table [x=x, y=y, col sep=comma] {figures/flowsInSequence/dwtcp/goodput_mbps-2.csv};
                \addplot[color=magenta, mark=x, line width=2.0pt, each nth point=3] table [x=x, y=y, col sep=comma] {figures/flowsInSequence/dwtcp/goodput_mbps-3.csv};
                \addplot[color=black, mark=x, line width=2.0pt, each nth point=3] table [x=x, y=y, col sep=comma] {figures/flowsInSequence/dwtcp/goodput_mbps-4.csv};
            \end{axis}
        \end{tikzpicture}
        \caption{DWTCP}
        \label{FIG:performance_overview_DWTCP}
    \end{subfigure}
    \caption{Long flows joining (and leaving) in sequence. \proto maintains small queues and achieves high fairness. Besides, \proto converges faster, maintains fairness, and gives away bandwidth to new flows quickly.}
    \label{FIG:performance_overview_5flows}
\end{figure*}
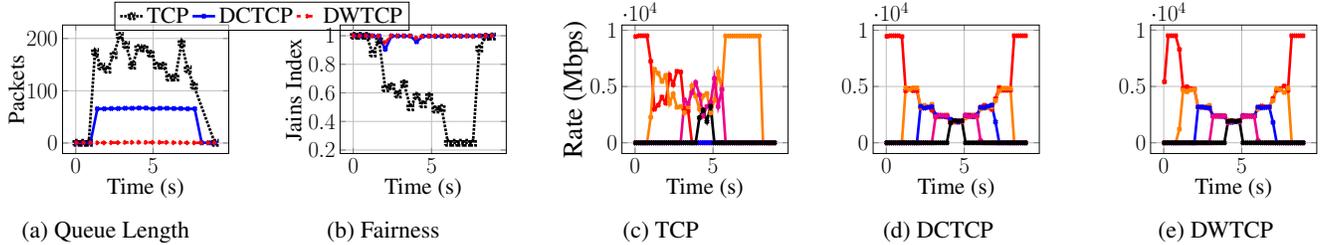

\section{Evaluation}
\label{sec:evaluation}
We evaluate \proto using a combination of small scale real testbed experiments and large scale NS3~\cite{ns3} simulations.
We take a closer look at \proto performance in the following aspects.

\noindent
\paragraph{\textbf{How does \proto perform in practice?}}
We use a simple experiment with the combination of short and long flows to demonstrate that \service can help \proto achieve:
i) near-zero queues, 
ii) small flow completion times (improvement by up to 2x compared to DCTCP, 4x compared to TCP), and 
iii) high throughput comparable to existing protocols ($1-3\%$ reduction)~\cite{alizadeh2010data,allman1999tcp}. Maintaining high throughput with near-zero queues is a challenging task for TCP algorithms. \proto, despite the \service overhead and near-zero queues, is able to achieve similar throughput as TCP and DCTCP.
We observe similar behaviors in testbed and NS3 simulations.

\noindent
\paragraph{\textbf{How does \proto perform in large scale datacenters?}}
Using large scale NS3 simulations, we demonstrate that \proto, compared to DCTCP~\cite{alizadeh2010data} and HPCC~\cite{ben2020pint}, improves latency by up to $2x$ at the mean and up to $3.5x$ at the tail with the websearch workload~\cite{alizadeh2010data} and up to $1.5x$ at the mean and up to $3x$ at the tail with the datamining workload.
%
HPCC performs better (up to 31\%) than \proto at low loads.

\noindent
\paragraph{\textbf{What is the impact of individual parameters on \proto?}}
We demonstrate that \proto is not very sensitive to control parameters ($\alpha$ and $\beta$), and it can work across a range of scenarios such as varying number of senders and long haul networks.



\subsection{Simulation Setup}
We start with NS3~\cite{ns3} simulations for evaluating \proto. We present testbed experiments in the following section.

\paragraph{\textbf{Topology.}}
We consider two different topologies: a 144-node real data center topology and a single bottleneck topology.
For the leaf-spine topology, edge links are $10$~Gbps, core links are $40$~Gbps, RTT is $100 \mu s$ and switch buffer size is set to  $250$ packets. 
Also, we consider a single bottleneck topology with RTT of $100 \mu s$  to keep it consistent with our real testbed setup.

\paragraph{\textbf{Traffic Patterns.}} We generate traffic using the websearch \cite{alizadeh2010data} and datamining~\cite{greenberg2009vl2} workloads. 
All-to-all traffic pattern with poisson arrival process and an average load of $20\% - 80\%$ of link capacity is used.
%

\paragraph{\textbf{Comparisons and Metrics.}} We evaluate and compare \proto to DCTCP, TCP NewReno (noted TCP in the figures), HPCC-PINT (noted HPCC in the figures). 
We report flow completion times (slowdown), fairness (Jains Index~\cite{JainIndex}), throughput and queue size. 
Using the publicly available NS3 simulator for HPCC, we also evaluate Timely~\cite{Timely} and DCQCN~\cite{DCQCN}, and our results confirm what shown in~\cite{li2019hpcc}, that is, HPCC significantly outperforms both Timely and DCQCN. As a result, in the interest of space, we only include the results of HPCC in our evaluation figures.
While the implementation of SWIFT~\cite{swift2020} is not publicly available for us to experimentally compare with \proto, we believe the amplification in delay and early notification through \service can lead to better or comparable results without requiring SWIFT's NIC hardware timestamps.
%

\paragraph{\textbf{\proto Settings.}}
For \proto, we set the low priority queue size in switches as $10$~\service packets ($\approx 600bytes$).
\proto sends $1$~\service packet of size $64B$ each RTT, unless specified otherwise. 
It uses $\beta$ of $0.001$, $K=100$ and initial value of $\alpha=20$ (equivalent of 1MTU). We also evaluate the impact of varying these parameters in \S\ref{SEC:deepdive}.

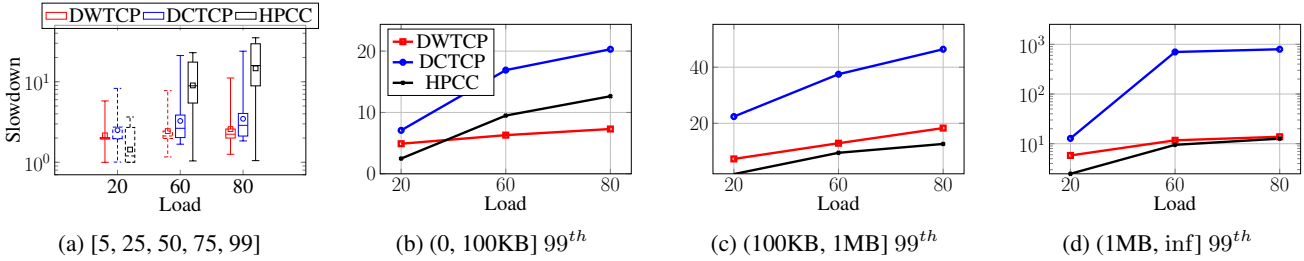
\begin{figure*}[ht!]
    \centering
    \begin{subfigure}[b]{0.24\textwidth}
        \centering

       \begin{tikzpicture}[scale=0.45, font=\LARGE]
        \begin{axis} [area legend,legend columns=-1, legend style={at={(0.5,1.15)},anchor=north,legend cell align=left, fill opacity=0.6, text opacity=1},
        height=6.0cm, width=9cm,
        xmin=0, xmax=25, 
          xtick={1,2,3},
          xticklabels={20, 60, 80},
          xmin=0,xmax=4, xlabel=Load, ylabel= Slowdown, y label style={at={(0.0,0.5)}}, ymode=log, boxplot/draw direction=y,
          /pgfplots/boxplot/box extend=0.15
          ]


          
        \addplot+ [color=red,thick,mark=square,
        boxplot prepared={
          median=2.231098,
          upper quartile=2.590517,
          lower quartile=2.0005161,
          upper whisker=11.13233,
          lower whisker=1.25922805,
          draw position=2.8
        },
        ] coordinates {(2.8,2.5839290)};
        \addlegendentry{DWTCP}

        \addplot+ [color=blue,thick, mark=o,
        boxplot prepared={
          median=2.8766,
          upper quartile=4.0414111,
          lower quartile=2.118240,
          upper whisker=23.9635,
          lower whisker=1.841414,
          draw position=3.0
        },
        ] coordinates {(3.0,3.468008)};
        \addlegendentry{DCTCP}

        \addplot+ [color=black,thick,mark=square,
        boxplot prepared={
          median=15.9,
          upper quartile=29.590517,
          lower quartile=8.89,
          upper whisker=35.13233,
          lower whisker=1.049,
          draw position=3.2
        },
        ] coordinates {(3.2,14.66)};
        \addlegendentry{HPCC}

        
       \addplot+ [color=black,thick,mark=square,
       boxplot prepared={
         median=8.9,
         upper quartile=17.517,
         lower quartile=5.44,
         upper whisker=22.93233,
         lower whisker=1.04,
         draw position=2.2
       },
       ] coordinates {(2.2,9)};

       \addplot+ [color=blue, thick, mark=o,
          boxplot prepared={
            median=2.6435,
            upper quartile=3.8588,
            lower quartile=2.027457,
            upper whisker=21.14611,
            lower whisker=1.678107,
            draw position=2.0
          },
          ] coordinates {(2.0,3.26150)};

          \addplot+ [color=red,mark=square, thick, 
          boxplot prepared={
            median=2.123,
            upper quartile=2.4424,
            lower quartile=1.975687,
            upper whisker=7.76641,
            lower whisker=1.1650444,
            draw position=1.8
          },
          ] coordinates {(1.8,2.4399)};

        
          \addplot+ [color=black,thick,mark=square,
          boxplot prepared={
            median=1.2,
            upper quartile=2.71,
            lower quartile=1.001,
            upper whisker=3.65233,
            lower whisker=1.0,
            draw position=1.2
          },
          ] coordinates {(1.2,1.44)};

            \addplot+ [color=blue,mark=o,
            boxplot prepared={
              median=1.970,
              upper quartile=2.73375,
              lower quartile=1.947547,
              upper whisker=8.2746714,
              lower whisker=1.00685246,
              draw position=1.0
            },
            ] coordinates {(1.0, 2.50561653)};
  
            \addplot+ [color=red,mark=square, thick, solid,
            boxplot prepared={
              median=1.9637,
              upper quartile=2.0314,
              lower quartile=1.94562,
              upper whisker=5.78045,
              lower whisker=1.0000657,
              draw position=0.8
            },
            ] coordinates {(0.8,2.16224536)};

        \end{axis}
    \end{tikzpicture}
    \caption{[5, 25, 50, 75, 99]} 
    \label{FIG:datamining_a}
    \end{subfigure}
    \hspace*{\fill}
    \begin{subfigure}[b]{0.24\textwidth}
        \centering
        \begin{tikzpicture}[scale=0.45, font=\LARGE]
            \begin{axis}[
                ymin=0,
                width=9cm,
                height=6cm,
                xlabel=Load,
                symbolic x coords={$20$, $60$, $80$},
                xtick = data,
                enlarge y limits={value=0.2,upper},
                legend pos=north west,
                ymajorgrids=true,xmajorgrids=true,
            ]

            \addplot[color=red, mark=square, line width=2.0pt] coordinates {($20$, 4.9) ($60$, 6.3) ($80$, 7.3)};
            \addplot[color=blue, mark=o, line width=2.0pt] coordinates {($20$,  7.08) ($60$, 16.9) ($80$, 20.3)};
            \addplot[color=black, mark=x, line width=2.0pt] coordinates {($20$,  2.48) ($60$, 9.49) ($80$, 12.64)};
            \legend{DWTCP, DCTCP, HPCC}
            \end{axis}
        \end{tikzpicture}
        \caption{(0, 100KB] $99^{th}$} 
        \label{FIG:datamining_b}
    \end{subfigure}
    \hspace*{\fill}
    \begin{subfigure}[b]{0.24\textwidth}
        \centering
        \begin{tikzpicture}[scale=0.45, font=\LARGE]
            \begin{axis}[
                ymin=2,
                width=9cm,
                height=6cm,
                xlabel=Load,
                symbolic x coords={$20$, $60$, $80$},
                xtick = data,
                enlarge y limits={value=0.2,upper},
                legend pos=north west,
                ymajorgrids=true,xmajorgrids=true,
            ]

            \addplot[color=red, mark=square, line width=2.0pt] coordinates {($20$, 7.3) ($60$, 12.9) ($80$, 18.3)};
            \addplot[color=blue, mark=o, line width=2.0pt] coordinates {($20$,  22.4) ($60$, 37.5) ($80$, 46.4)};
            \addplot[color=black, mark=x, line width=2.0pt] coordinates {($20$,  1.9) ($60$, 9.49) ($80$, 12.64)};
            \end{axis}
        \end{tikzpicture}
        \caption{(100KB, 1MB] $99^{th}$} 
        \label{FIG:datamining_c}
    \end{subfigure}
    \hspace*{\fill}
    \begin{subfigure}[b]{0.24\textwidth}
        \centering
        \begin{tikzpicture}[scale=0.45, font=\LARGE]
            \begin{axis}[
                width=9cm,
                height=6cm,
                xlabel=Load,
                ymode=log,
                symbolic x coords={$20$, $60$, $80$},
                xtick = data,
                enlarge y limits={value=0.2,upper},
                legend pos=north west,
                ymajorgrids=true,xmajorgrids=true,
            ]

            \addplot[color=red, mark=square, line width=2.0pt] coordinates {($20$, 5.8) ($60$, 11.7) ($80$, 13.8)};
            \addplot[color=blue, mark=o, line width=2.0pt] coordinates {($20$,  12.8) ($60$,  700.8) ($80$, 795.7)};
            \addplot[color=black, mark=x, line width=2.0pt] coordinates {($20$,  2.48) ($60$, 9.49) ($80$, 12.64)};
            \end{axis}
        \end{tikzpicture}
        \caption{(1MB, $\inf$] $99^{th}$} 
        \label{FIG:datamining_d}
    \end{subfigure}

        \caption{Datamining workload: a) Percentile plots. Small circle and square show the mean slowdown. Compared to DCTCP (and HPCC), \proto improves up to 1x (2.4x) mean slowdown and up to 3x (3.2x) for the tail slowdown. (b-d) \proto significantly improves tail slowdown for small by 3x, medium by 2.4x and large flows by 100x compared to DCTCP. \proto improves slowdown by 1.7x for small flows and achieves similar performance for medium and long flows compared to HPCC.}
        \label{FIG:datamining}
        \vspace{-0.1in}
\end{figure*}

\subsection{Performance Overview}
Figure~\ref{FIG:motivation} shows a glimpse of \proto in maintaining small queues, achieving fast convergence, high throughput and small flow completion times. 
To highlight the performance of \proto in more dynamic scenarios, we consider a single bottleneck topology and generate five flows between different senders and receivers. Flows start $1s$ apart, and once we have five active flows, they leave $1s$ apart.

\paragraph{\textbf{\proto is Stable and Achieves Fairness.}}
Figure~\ref{FIG:performance_overview_5flows} shows that \proto is very stable in maintaining small (near-zero) queues (Figure~\ref{FIG:poql}) and achieves high fairness (Figure~\ref{FIG:pof}) compared to TCP (NewReno), 
 and DCTCP.
\proto reacts quickly to congestion and adapts its rate dynamically to maintain stable queue sizes.
DCTCP is also very stable in maintaining queue sizes. However, it maintains queue levels around the threshold (65 in this case), which can impact the latency of short flows. 
This stable behavior of \proto helps achieve high fairness among the flows (Figure~\ref{FIG:pof}). As shown, the fairness index for \proto is very close to 1 (i.e., optimal fairness index). When a new flow arrives (or leaves), other flows quickly give up (or utilize) the link capacity.

\paragraph{\textbf{\proto Reacts Rapidly to Congestion.}}
Figures~\ref{FIG:performance_overview_TCP}, 
\ref{FIG:performance_overview_DCTCP}, and 
\ref{FIG:performance_overview_DWTCP} depicts the throughput of TCP, 
DCTCP and \proto respectively.
We notice that \proto reacts quickly to the congestion from the new flows and converges to fair share without any peak oscillations, unlike other protocols. Furthermore, due to dynamic bandwidth availability estimation, it quickly uses up any spare bandwidth when the flows leave.
Note that, although TCP and DCTCP maintain higher queues, \proto is able to maintain similar throughput even with near-zero queues as shown in Figure~\ref{FIG:poql} due to its fast convergence properties.

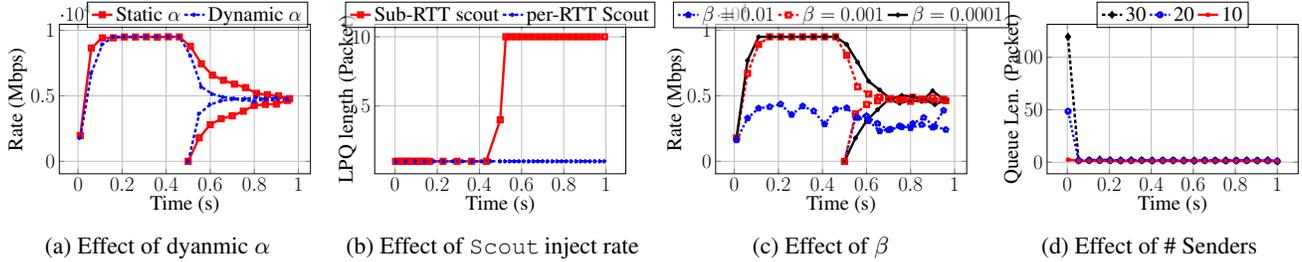
\begin{figure*}[htbp]
    \begin{subfigure}[b]{0.24\textwidth}
        \centering
        \begin{tikzpicture} [scale=0.45,font=\LARGE]
            \begin{axis}[legend columns=-1, legend pos=north east, 
            legend style={at={(0.55,1.15)},anchor=north,legend cell align=left, fill opacity=0.6, text opacity=1},
            xlabel=Time (s), 
            ylabel=Rate (Mbps), y label style={at={(0.0,0.5)}},
            height=6.cm, width=9cm,
            ymajorgrids=true,xmajorgrids=true,]
                \addplot[color=red, mark=square, line width=2.0pt, each nth point=5] table [x=x, y=y, col sep=comma] {figures/scout_parameters/alpha/goodput-0_static.csv};\addlegendentry{Static $\alpha$}   
                \addplot[color=blue, mark=x, densely dashed, line width=2.0pt, each nth point=5] table [x=x, y=y, col sep=comma] {figures/scout_parameters/alpha/goodput-0_dynamic.csv};\addlegendentry{Dynamic $\alpha$}
                \addplot[color=red, mark=square, line width=2.0pt, each nth point=5, domain=.5:1] table [x=x, y=y, col sep=comma] {figures/scout_parameters/alpha/goodput-1_static.csv};
                \addplot[color=blue, mark=x, densely dashed, line width=2.0pt, each nth point=5, domain=.5:1] table [x=x, y=y, col sep=comma] {figures/scout_parameters/alpha/goodput-1_dynamic.csv};
            \end{axis}
        \end{tikzpicture}
        \caption{Effect of dyanmic $\alpha$}
        \label{FIG:alpha_effect_convegence}
    \end{subfigure}
    \hspace*{\fill}
    \hspace*{\fill}
        \begin{subfigure}[b]{0.24\textwidth}
            \centering
            \begin{tikzpicture} [scale=0.45,font=\LARGE]
                \begin{axis}[legend columns=-1, legend style={at={(0.5,1.15)},anchor=north,legend cell align=left, fill opacity=0.6, text opacity=1},
                xlabel=Time (s),
                ylabel=LPQ length (Packet), y label style={at={(0.06,0.5)}},
                height=6.cm, width=9cm,
                ymajorgrids=true,xmajorgrids=true,]
                \addplot[color=red, mark=square, line width=2.0pt, each nth point=3] table [x=x, y=y, col sep=comma] {figures/scout_parameters/alpha/LPQ_static.csv};\addlegendentry{Sub-RTT scout}   
                \addplot[color=blue, mark=x, densely dashed, line width=2.0pt, each nth point=3] table [x=x, y=y, col sep=comma] {figures/scout_parameters/alpha/LPQ_dynamic.csv};\addlegendentry{per-RTT Scout}
            \end{axis}
        \end{tikzpicture}
        \caption{Effect of \service inject rate}
        \label{FIG:alpha_effect_LPQ}
        \end{subfigure}
    \hspace*{\fill}
        \begin{subfigure}[b]{0.24\textwidth}
            \centering
            \begin{tikzpicture} [scale=0.45,font=\LARGE]
                \begin{axis}[legend columns=-1, legend pos=north east, legend style={at={(0.5,1.15)},anchor=north, legend cell align=left, fill opacity=0.6, text opacity=1},
                xlabel=Time (s), height=6.cm, width=9cm,
                ylabel=Rate (Mbps), y label style={at={(0.01,0.5)}},
                ymajorgrids=true,xmajorgrids=true,]
                \addplot[color=blue, mark=pentagon,mark options=solid, dotted, line width=2.0pt, each nth point=5] table [x=x, y=y, col sep=comma] {figures/scout_parameters/beta/goodput-1_B.01.csv};
                \addlegendentry{$\beta=0.01$}
                \addplot[color=red, mark=square, densely dashed,mark options=solid, line width=2.0pt, each nth point=5] table [x=x, y=y, col sep=comma] {figures/scout_parameters/beta/goodput-0_B.001.csv};
                \addlegendentry{$\beta=0.001$}
                \addplot[color=black, mark=+, line width=2.0pt, each nth point=5] table [x=x, y=y, col sep=comma] {figures/scout_parameters/beta/goodput-0_B.0001.csv};
                \addlegendentry{$\beta=0.0001$}

                \addplot[color=black, mark=+, line width=2.0pt, each nth point=5] table [x=x, y=y, col sep=comma] {figures/scout_parameters/beta/goodput-1_B.0001.csv};
                \addplot[color=red, mark=square, densely dashed,mark options=solid, line width=2.0pt, each nth point=5] table [x=x, y=y, col sep=comma] {figures/scout_parameters/beta/goodput-1_B.001.csv};
                \addplot[color=blue, mark=pentagon,mark options=solid, dotted, line width=2.0pt, each nth point=5] table [x=x, y=y, col sep=comma] {figures/scout_parameters/beta/goodput-0_B.01.csv};
            \end{axis}
            \end{tikzpicture}
            \caption{Effect of $\beta$}
            \label{FIG:beta_effect_convegence}
        \end{subfigure}
        \hspace*{\fill}
        \begin{subfigure}[b]{0.24\linewidth}
            \centering
            \begin{tikzpicture}[scale=0.45,font=\LARGE]
                \begin{axis}[legend style={at={(0.5,1.15)}, legend columns=-1, anchor=north,legend cell align=left, fill opacity=0.6, text opacity=1},
                    xlabel=Time (s), ylabel=Queue Len. (Packet), y label style={at={(0.03,0.5)}},
                    , height=6.cm, width=9cm,
                    ymajorgrids=true,xmajorgrids=true,]
                    \addplot[color=black, mark=diamond, densely dashed,mark options=solid, line width=2.0pt, each nth point=5] table [x=x, y=y, col sep=comma] {figures/varying_sendersnum/30senders/packetsInQueue-r0-q0.csv};
                    \addlegendentry{30}
                    \addplot[color=blue, mark=o, dotted,mark options=solid, line width=2.0pt, each nth point=5] table [x=x, y=y, col sep=comma] {figures/varying_sendersnum/20senders/packetsInQueue-r0-q0.csv};
                    \addlegendentry{20}
                    \addplot[color=red, mark=x, line width=2.0pt, each nth point=5] table [x=x, y=y, col sep=comma] {figures/varying_sendersnum/10senders/packetsInQueue-r0-q0.csv};
                    \addlegendentry{10}
                \end{axis}
            \end{tikzpicture}
            \caption{Effect of \# Senders}
            \label{FIG:varying_senders_num_queuesize}
        \end{subfigure}
        \vspace{-0.1in}
    \caption{\service optimizations significantly improve the goodput convergence, fairness, reduce LPQ size, and make it stable across different settings.}
    \label{FIG:scout_parameters_effects}
    \vspace{-0.2in}
\end{figure*}

\subsection{Benchmark Workloads}
\label{sec:benchmark}
Next, we evaluate \proto in more realistic datacenter settings using all-to-all traffic pattern for the websearch~\cite{alizadeh2010data} and datamining~\cite{greenberg2009vl2} workloads in a typical leaf-spine topology and compare with DCTCP and HPCC.
%

\paragraph{\textbf{Datamining Workload.}}
Figures~\ref{FIG:datamining_a}, \ref{FIG:datamining_b}, \ref{FIG:datamining_c}, and \ref{FIG:datamining_d} show the percentile, the tail slowdown across small flows (0, 100KB], medium flows (100KB, 1MB], and large flows (10MB, $\inf$), respectively for the datamining workload. 
\proto outperforms DCTCP in both the average and 99th percentile slowdown. It improves the mean by up to 97\%, and tail up to 3x.
Moreover, \proto improves the tail slowdown by 1.5x-2.8x for the small flows, up to 2x-3.2x for the medium flows, and up to 3x-100x for the large flows. \proto improves the tail slowdown by up to 100x as it does not experience any timeouts at higher loads.
HPCC performs better (up to 31\%) at low loads due to flows starting at the line rate and faster convergence. However, \proto performs significantly better at higher loads and improves mean and tail slowdowns by up to 2x.
\proto maintains smaller queues, even at very high loads, as a result the tail completion times of the short flows improve by up to 1.7x, whereas the medium and long flows achieve similar slowdown as in HPCC.
Note that, more than 80\% of the flows in datamining workload belong to small flows.

\paragraph{\textbf{Websearch Workload.}}
We observe similar performance for the websearch workload. 
Figure~\ref{FIG:websearch} (see Appendix~\ref{sec:results}) shows that compared to DCTCP (and HPCC), \proto improves up to 2x (3x) mean slowdown and up to 3.5x (5x) for the tail slowdown. (b-d) DWTCP significantly
improves tail slowdown for small by 8x (3.7x), medium by 3.5x (10x) and large flows by 3.5x (15x).
Note that, compared to HPCC \proto performs better for both the short and long flows at high loads.

\subsection{\proto Deep-dive}\label{SEC:deepdive}
In this section, we study the impact of \proto parameters and varying network conditions, using a dumbbell topology.

\paragraph{\textbf{Effect of \service Coefficient ($\alpha$).}}
In this setup, we generate two long-lived flows. Flow 1 starts at time 0, and Flow 2 starts at time 0.5 seconds.
We compare the dynamic \service coefficient $\alpha$ with the fixed static value of 10.
Figure~\ref{FIG:alpha_effect_convegence} shows that dynamic $\alpha$ helps converge faster and improves fairness compared to a static $\alpha$.

\paragraph{\textbf{Effect of \service Injection Rate (x).}}
In the above experiment, for the static $\alpha$, we vary the \service injection rate instead of changing the \service coefficient. This mimics having a sub-RTT \service injection and also the static $\alpha$.
We can observe that sub-RTT \service injection rate significantly increases the \service overhead both in terms of network bandwidth utilization and the \service queue size in the switches, as shown in Figure~\ref{FIG:alpha_effect_LPQ}.
Therefore, the optimization of injecting one \service per RTT provides a very good guarantee that as we increase the number of flows, \proto can achieve high throughput and faster convergence without requiring larger \service buffers in the switches.

\paragraph{\textbf{Effect of $\beta$.}}
We repeat the above experiment settings with dynamic \service window and varying $\beta$.
$\beta$ factor controls the aggressiveness of the decrease, and setting it correctly greatly improves the convergence and fairness.
Figure~\ref{FIG:beta_effect_convegence} shows that using a very aggressive $\beta$ results in severe under-utilization. Similarly, setting it too small can increase the convergence time as the flows take longer to free up the bandwidth. This also results in larger queue sizes.

\paragraph{\textbf{Effect of Varying Number of Senders.}}
In this experiment, we increase the number of senders (receivers) to 10, 20, and 30 using default \proto parameters.
We can verify that \proto can adapt to different number of flows and can maintain smaller queues (Figure~\ref{FIG:varying_senders_num_queuesize}), achieve higher throughput and fairness (see Appendix~\ref{sec:results}, Figure~\ref{FIG:varyinge_senders_num}).
We observe an increase in the queue size in the beginning, which is the result of synchronization among the flows.

\paragraph{\textbf{Effect of Varying Network Conditions.}}
To observe the adaptability of \service to varying network conditions, we vary the link bandwidth every $0.1s$ as [0.4, 0.7, 1, 10, 0.8]~Gbps.
We run five long flows to observe the queue size, throughput, and fairness.
We observe that \proto quickly grabs any available bandwidth when the available bandwidth changes from 1~Gbps to 10~Gbps and it quickly converges to full link utilization, maintains small queues and achieves fairness. See Appendix~\ref{sec:results}, Figure~\ref{FIG:varying_bw} for more details.

\section{Testbed based Evaluation}
\label{sec:testbed} 
\subsection{Implementation}
The Linux implementation of \proto consists of three main components:

\paragraph{\textbf{Scout Generator.}}
In order to inject the \service packets, we initiate a \gls{udp} stream from each sender towards its destination using
the default installed packet generator on Linux (\codestyle{pktgen V2.75}). 
This packet generator is able to generate packets with inter-departure times as small as $7 \mu s$. 
Even though it sometimes experiences some jitters (with delays as large as $3\mu s$), this accuracy has shown to 
be enough for the range of \gls{rtt}s that we obtain in our testbed ($80 \mu s$).
Before sending each packet, in addition to marking the IP packets with
low priority DSCP values, it also fills the packet payload with their send time (note that there is a gap between
the marking time and the time that the packet is being placed on the wire).
In our setup we use microsecond clocks and a per datapath stream of \service packets rather than a per flow connection for 
the testbed experiments. Later, we show the difference of having a single stream of \service packets per datapath with the case that they are being initiated per flow. 

\paragraph{\textbf{Scout Reflector.}}
The reflector is a set of simple \codestyle{iptables} rules installed on the hosts which result in forwarding the 
incoming \service packets to their origin. This design in combination with using \codestyle{pktgen} allows us to implement a
prototype without touching the kernel code and also without compromising the performance.

\paragraph{\textbf{Congestion Algorithm.}} 
We implemented the \proto congestion algorithm in Linux (\codestyle{V5.4.0}) as a loadable kernel module. The \service analyzer is
currently embedded in this module via an added \codestyle{Netfilter} hook that listens for \gls{udp} packets. Upon receiving 
any relevant packet, we decode its sending timestamp to perform our delay calculation.   

\vspace{-0.1in}
\subsection{Testbed Description}
\paragraph{\textbf{Topology Setup.}}
We use a dumbbell topology in our testbed. There are 5 servers connected to each side
of this topology (i.e., 10 servers in total) with an access link of 10Gbps each. The bottleneck link is limited to 10Gbps and
300KB switch buffers. 
Each server is configured with a 64-core Intel(R) Xeon(R) Gold 5218 CPU \@ 2.30GHz, and two arrays of 32GB DRAM, and for the 
internal queueing discipline they use Linux's \codestyle{pfifo\_qdisc} (packet FIFO queueing discipline).
The bottleneck endpoint switches are set to use a strict priority for routing the traffic. 

\paragraph{\textbf{\proto Settings.}}
We impose priorities by setting the \codestyle{IPTOS} 
field of each \service packet to a lower priority than the normal TCP traffic. 
We use a fixed \service window size per datapath. Each \gls{udp} \service packet is of size 64B, and they leave 
their source every $7\mu s$. The lower priority queue of the switch was also set to 640B (i.e., 10 packets).
Similar to simulation setup, we set $\beta$ to 0.001, and $K$ to 100. 
\paragraph{\textbf{Performance Metrics.}}
We use \gls{cwnd}, \gls{rtt}, goodput of long flows and flow completion times (FCT) of the short flows as performance metrics.
We assume that there is enough traffic being generated to congest the bottleneck (unless stated otherwise). Throughout each experiment, 
we periodically poll (average over buckets of $20ms$) all the connection metrics of interest using a dedicated core on each machine. 

\subsection{Performance Overview}
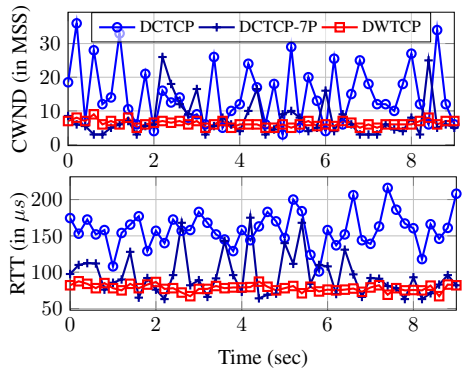
\begin{figure}[t]
    \centering
    \begin{subfigure}[b]{1\linewidth}
        \centering
        \begin{tikzpicture}[scale=0.8]
            \begin{axis}[
                legend style={at={(0.5,1)}, legend columns=-1,nodes={scale=0.8, transform shape},anchor=north,legend cell align=left, fill opacity=0.6, text opacity=1},
                ylabel=CWND (in MSS), y label style={at={(0.01,0.5)}}, height=3.8cm, width=8cm,
                ymajorgrids=true,xmajorgrids=true, xmin=0, xmax=9, 
                y label style={at={(0.07,0.5)}},
                ]
                \addplot[color=blue, mark=o, line width=1pt, each nth point=2] table [x=timestamp, y=cwnd, col sep=comma] {figures/TESTBED/M-steady/cwnd-DCTCP.csv};
                \addlegendentry{DCTCP}
                \addplot[color=blue!60!black, mark=+, line width=1pt, each nth point=2] table [x=timestamp, y=cwnd, col sep=comma] {figures/TESTBED/M-steady/cwnd-DCTCP-7P.csv};
                \addlegendentry{DCTCP-7P}
                \addplot[color=red, mark=square, line width=1pt, each nth point=2] table [x=timestamp, y=cwnd, col sep=comma] {figures/TESTBED/M-steady/cwnd-DWTCP.csv};
                \addlegendentry{DWTCP}
            \end{axis}
        \end{tikzpicture}
        \label{FIG:testbed-steady-cwnd}
    \end{subfigure}

    \begin{subfigure}[b]{1\linewidth}
        \centering
        \begin{tikzpicture}[scale=0.8]
            \begin{axis}[
                xlabel=Time (sec), ylabel=RTT (in $\mu s$), ylabel style={at={(0.01,0.5)}}, height=3.8cm, width=8cm,
                ymajorgrids=true,xmajorgrids=true, xmin=0, xmax=9, 
                y label style={at={(0.06,0.5)}},
                ]
                \addplot[color=blue, mark=o, line width=1pt, each nth point=2] table [x=timestamp, y=rtt, col sep=comma] {figures/TESTBED/M-steady/rtt-DCTCP.csv};
                \addlegendentry{DCTCP}
                \addplot[color=blue!60!black, mark=+, line width=1pt, each nth point=2] table [x=timestamp, y=rtt, col sep=comma] {figures/TESTBED/M-steady/rtt-DCTCP-7P.csv};
                \addlegendentry{DCTCP-7P}
                \addplot[color=red, mark=square, line width=1pt, each nth point=2] table [x=timestamp, y=rtt, col sep=comma] {figures/TESTBED/M-steady/rtt-DWTCP.csv};
                \addlegendentry{DWTCP}
                \legend{};
            \end{axis}
        \end{tikzpicture}
        \label{FIG:testbed-steady-rtt}
    \end{subfigure}
    \vspace{-0.3in}
    \caption{Steady state test with 10 flows. \proto is stable and maintains smaller queues. \gls{dctcp} with smaller threshold (K) experiences underutilization.}
    \label{FIG:testbed-steady}
    \vspace{-0.2in}
\end{figure}

\paragraph{\textbf{Steady State.}}
In this experiment we use five hosts as the senders, each with two flows that start simultaneously (i.e., 10 flows).
For \gls{dctcp}, we consider an additional setting than the originally suggested parameters for its ECN marking thresholds.  
In this setting, we set the ECN threshold marking to be 7 packets rather than the suggested 65 packets 
for 10Gbps bottlenecks. This threshold is equivalent with DWTCP's queue sensitivity parameter $K$ which is set to 7 packets. 
We observe significant underutilization (around 15\%) in this case as one could expect from DCTCP's dynamic.
Figure~\ref{FIG:testbed-steady} compares the \gls{rtt} and \gls{cwnd} of one of these flows from each scheme. 
One can see that \proto achieves minimum \gls{rtt} and zero queue loss while maintaining link utilization close to the maximum. 
This is while \gls{dctcp} has 0.01\% packet retransmission on average in the steady case.

\begin{figure*}[t!]
    \centering
    \begin{subfigure}[t]{0.48\linewidth}
        \centering
        \includegraphics[width=0.9\linewidth]{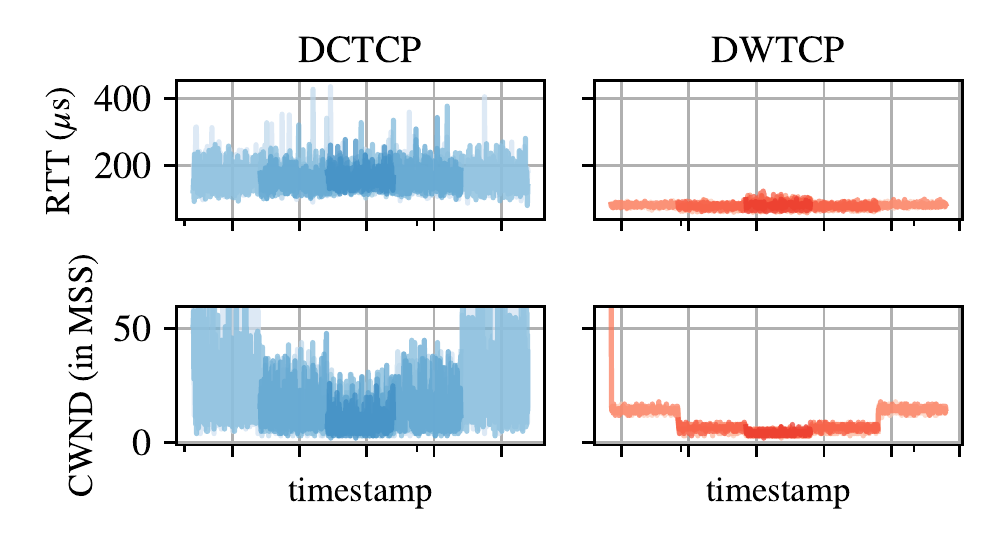}
        \caption{RTT ($\mu s$)}
        \label{FIG:testbed-incdec-series}
    \end{subfigure}
    \begin{subfigure}[t]{0.28\linewidth}
        \centering
        \begin{tikzpicture}[scale=0.55, font=\Large]
            \begin{axis}[
                height=7cm,
                width=8cm,
                ybar,
                ymin=0,
                ymax=300,
                bar width=8pt,
                ylabel=RTT ($\mu s$),y label style={at={(0.05,0.52)}},
                xlabel=Percentile,
                symbolic x coords={Mean, $50^{th}$, $90^{th}$, $99^{th}$, $99.9^{th}$},
                xtick = data,
                enlarge x limits=0.11,
                enlarge y limits={value=0.22,upper},
                legend pos=north west,
                ymajorgrids=true,xmajorgrids=true,
                legend columns=1,
                legend style={nodes={scale=0.8, transform shape},anchor=north west,legend cell align=left, fill opacity=0.6, text opacity=1},
            ]
            \addplot[pattern color=blue, pattern=crosshatch] coordinates {(Mean, 162) ($50^{th}$, 158) ($90^{th}$, 205) ($99^{th}$, 260) ($99.9^{th}$, 354)};
            \addplot[pattern color=red, pattern=north east lines] coordinates {(Mean, 80) ($50^{th}$, 81) ($90^{th}$, 90) ($99^{th}$, 101) ($99.9^{th}$, 113)};
            \addplot[fill=red] coordinates {(Mean, 84) ($50^{th}$, 80) ($90^{th}$, 111) ($99^{th}$, 141) ($99.9^{th}$, 166)};
            \legend{DCTCP, DWTCP, DWTCP-FLOW}
            \end{axis}
        \end{tikzpicture}
        \caption{RTT Tail}
        \label{FIG:testbed-incdec-rtt}
    \end{subfigure}
    \begin{subfigure}[t]{0.22\linewidth}
        \begin{tikzpicture}[scale=0.55, font=\Large]
            \begin{axis}[
                height=7cm,
                width=8cm,
                ylabel=CDF, y label style={at={(0.1,0.92)}},
                xlabel=Goodput (Gbps),
                ymajorgrids=true,xmajorgrids=true,
                legend style={nodes={scale=0.8, transform shape}, 
                              legend pos=north west,
                              legend cell align=right, 
                              fill opacity=0.6, 
                              text opacity=1, 
                              legend columns=1,
                              }
            ]   
            \addplot[color=blue, loosely dashed, line width=2.0pt] table [x=r, y=p, col sep=comma] {figures/TESTBED/M-inc-dec/cdf-dctcp.csv};
            \addlegendentry{DCTCP}
            \addplot[color=red, dotted, line width=2.0pt] table [x=r, y=p, col sep=comma] {figures/TESTBED/M-inc-dec/cdf-dwtcp-host.csv};
            \addlegendentry{DWTCP}
            \addplot[color=red, line width=2.0pt] table [x=r, y=p, col sep=comma] {figures/TESTBED/M-inc-dec/cdf-dwtcp.csv};
            \addlegendentry{DWTCP-FLOW}

        \end{axis}
        \end{tikzpicture}
        \caption{Goodput CDF}
        \label{FIG:testbed-incdec-cdf}
    \end{subfigure}
    \vspace{-0.1in}
    \caption{Long flows joining (and leaving) in sequence. \proto maintains small queues and achieves high fairness. Besides, \proto converges faster, maintains fairness, and gives away bandwidth to new flows quickly}
    \label{FIG:testbed-incdec}
    \vspace{-0.1in}
\end{figure*}
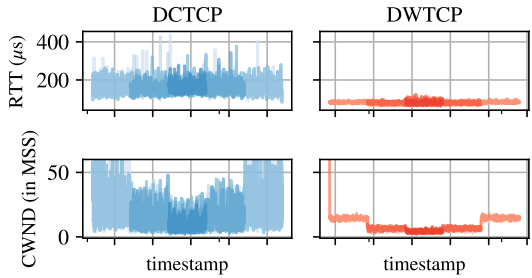
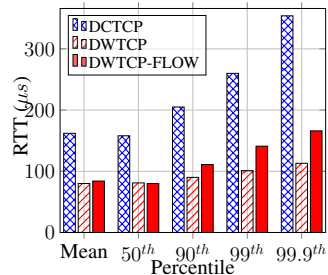
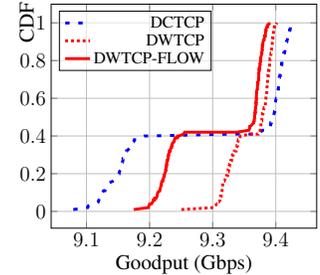

\paragraph{\textbf{Synchronized Flows.}}
In order to measure the responsiveness of \proto in sudden network changes, we initiate three batches of five simultaneous flows with the contribution of one flow from each sender in a batch.
Each batch starts and stops at chunks of five seconds.
The small amplitude of the changes in the CWND and RTT (Figure~\ref{FIG:testbed-incdec-series}) for 
the flows in \proto compared to DCTCP explains the near-perfect fairness that \proto achieves.
Figure~\ref{FIG:testbed-incdec-rtt} shows that while \proto maintains $2-3X$ faster
RTTs, it also reaches close to the maximum throughput as depicted by the goodput CDF in Figure~\ref{FIG:testbed-incdec-cdf}.
The responsiveness power of \service becomes more apparent by investigating the reason for the throughput loss in DCTCP. 
Figure~\ref{FIG:testbed-incdec-cdf} depicts that \gls{dwtcp} maintains on average higher goodput compared to DCTCP because DWTCP doesn't experience packet retransmission. In addition, DWTCP keeps the link utilization below the maximum link capacity. Such behavior is intended by design to enable DWTCP's smooth transition during sudden changes.

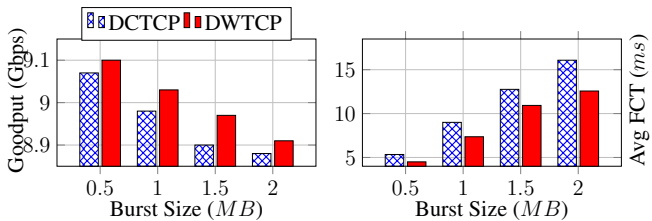
\begin{figure}[ht!]
    \centering

    \begin{subfigure}[b]{0.48\linewidth}
        \centering
        \begin{tikzpicture}[scale=0.7, font=\large]
            \begin{axis}[
                ybar,
                ymin=8.85,
                width=6.5cm,
                height=4cm,
                bar width=10pt,
                xlabel=Burst Size ($MB$),
                ylabel = {Goodput (Gbps)},y label style={at={(0.1,0.52)}},
                symbolic x coords={$0.5$,$1$,$1.5$,$2$},
                xtick = data,
                enlarge y limits={value=0.2,upper},
                enlarge x limits=0.25,
                ymajorgrids=true,xmajorgrids=true,
                legend style={at={(0.5,1.25)},legend columns=-1,anchor=north,legend cell align=left, fill=white, fill opacity=1, draw opacity=1, text opacity =1},
            ]
            \addplot[pattern color=blue, pattern=crosshatch]
            coordinates {
                ($0.5$,9.07) 
                ($1$,8.98) 
                ($1.5$,8.9) 
                ($2$,8.88)};
            \addplot[fill=red] 
            coordinates {
                ($0.5$,9.1) 
                ($1$,9.03) 
                ($1.5$,8.97) 
                ($2$,8.91)};
            \legend{DCTCP, DWTCP}
            \end{axis}
        \end{tikzpicture}
    \end{subfigure}
    \hspace*{\fill}
    \begin{subfigure}[b]{0.48\linewidth}
        \centering
        \begin{tikzpicture}[scale=0.7, font=\large]
            \begin{axis}[
                ybar,
                ymin=4,
                width=6.5cm,
                height=4cm,
                bar width=10pt,
                xlabel=Burst Size ($MB$),
                ylabel = {Avg FCT ($ms$)},y label style={at={(1.32,0.52)}},
                symbolic x coords={$0.5$,$1$,$1.5$,$2$},
                xtick = data,
                enlarge y limits={value=0.2,upper},
                enlarge x limits=0.25,
                ymajorgrids=true,xmajorgrids=true,
                legend style={at={(0.5,.99)},legend columns=-1,anchor=north,legend cell align=left, fill=white, fill opacity=0.6, draw opacity=1, text opacity =1},
            ]
            \addplot[pattern color=blue, pattern=crosshatch]
            coordinates {
                ($0.5$,5.34)
                ($1$,9.00)
                ($1.5$,12.76)
                ($2$,16.08)};
            \addplot[fill=red] 
            coordinates {
                ($0.5$,4.5) 
                ($1$,7.36)
                ($1.5$,10.93)
                ($2$,12.57)};
            \legend{}
            \end{axis}
        \end{tikzpicture}
    \end{subfigure}
    \caption{FCT of short flows and goodput of long flows.}
    \label{FIG:testbed_fct}
    \vspace{-0.2in}
\end{figure}

\paragraph{\textbf{Flow Completion Times.}}
In this experiment, we measure the impact of bursts.
We consider two background flows from two different senders to one client. 
And six other flows send  
sudden bursts of fixed size towards the same client. 
We repeat bursts every 500 milliseconds and vary their size from 0.5MB to 2MB per connection.
Figure~\ref{FIG:testbed_fct} shows the average flow completion times of these bursts 
in comparison with the goodput of the background flows.
We observe that \proto consistently obtains higher goodput than DCTCP at different burst sizes, while achieving smaller latency of $81\mu s$ ($163\mu s$ for DCTCP)
In addition, 
\proto improved the FCT of short flows by at least 15\%.

\section{Related Work and Discussion}
\vspace{-0.1in}
\noindent
\paragraph{\textbf{Comparison with Delay-based CC.}}
%
%
Delay based mechanisms, such as Timely~\cite{Timely} or SWIFT~\cite{swift2020}, experience stability issues or need to apply moving average to mitigate noise effect while calculating the delay at the host.
On the other hand, the arrival of \Service ACK is a clear indication of bandwidth availability. has a very robust signal for bandwidth availability as the arrival of the \service
Moreover, \service reacts based on the delay or loss of the \service traffic in low priority queue, which means it has a much faster response time in detecting congestion in the high priority queue. 
%
As a result, \service is more stable and does not require hardware timers (or averaging) to compute delays, as in SWIFT.

\noindent
\paragraph{\textbf{Comparison with Credit-based CC.}}
Credit-based congestion control algorithms often provide better performance than traditional congestion signals~\cite{cho2017credit,liu2004high}.
However, hop-by-hop credit-based mechanisms (such as Infiniband~\cite{liu2004high}) are expensive, difficult to manage, and have a high overhead. Some well known challenges of these mechanisms are: i) Interaction with congestion control protocol and ii) Deadlock issue.
Similarly,	path-based credit-based mechanisms need switch modifications, or reserve a fraction of bandwidth to manage and throttle credits~\cite{cho2017credit}. 
Compared to these mechanisms, \service is easier to implement, does not require any switch modifications, and only requires small buffers (less than 1MTU) for the \service traffic.

\noindent
\paragraph{\textbf{Comparison with Out-of-band OAM.}} 
Prior art on out-of-band OAM, such as PathChirp~\cite{pathChirp} and BART~\cite{bart}, aim to estimate available bandwidth per data path. However, these mechanisms rely on sending probe packets with the same priority as the normal traffic, which incurs significant overhead, and can impact the ongoing network traffic.
Moreover, for delay and available bandwidth measurements, such approaches are very noisy and depend on the state of the main data queue.
On the other hand, \service uses low priority traffic which incurs negligible overhead, does not impact the main traffic, and provides a much amplified and stable signal.

\noindent
\paragraph{\textbf{Interaction with Multipath Routing.}}
\service packets need to have the same hash value as the normal traffic to follow the same path, when ECMP~\cite{alizadeh2014conga} is enabled. 
To achieve this, in our current implementation, we modify the hashing algorithm at the switch based on the ToS field. 
An alternative approach is to use a zero payload packet, from the same connection,  or add an extra flag to the TCP header to denote \service packet.

\noindent
\paragraph{\textbf{Comparison with Flow Scheduling-based Schemes.}}
In the recent past, many scheduling based transport schemes have been proposed for data center networks~\cite{alizadeh2013pfabric,li2019hpcc,munir2014friends,bai2015information,munir2013minimizing, montazeri2018homa}.
These schemes aim at minimizing the FCTs of latency sensitive flows and achieve near optimal performance.
\proto is not a scheduling-based scheme, however, it maintains very small queues at the switches, which is very useful for the latency sensitive flows. 
In this paper, we demonstrate that \proto provides significant improvement, compared to DCTCP~\cite{alizadeh2010data}, in terms of FCTs. 
It will be interesting to evaluate how it compares to deployment friendly scheduling-based schemes~\cite{munir2013minimizing,munir2014friends, bai2015information}. We leave this as a future work.

\section{Conclusion and future work}
\vspace{-0.1in}
In this paper, we presented \service service, an early indicator of congestion and bandwidth availability based on sending probe traffic with low priority. We also presented \gls{dwtcp} as a Scout-based congestion control mechanism.
Our extensive simulations and testbed experiments show that \gls{dwtcp} outperforms TCP NewReno, and DCTCP in maintaining close to zero queue length, and reduces network latency to near optimal. In addition, \gls{dwtcp} outperforms HPCC for high load scenarios and achieves comparable results at low loads without the need for switch modifications.
The experimental results also show significant reductions in flow completion times (mean and tail) compared to DCTCP. We have mathematically proved that \gls{dwtcp} is stable (and the stability conditions are deduced), and has one stable equilibrium point.

The \service has a very nice property that as it provides the signal several RTTs earlier, it makes a congestion control protocol less sensitive to the timestamping delays and therefore it eliminates the need for hardware timestamping even for small values of RTTs.
Even though our experiments show \service has reasonable overhead, one can imagine the overhead can be further reduced by using a data packet, or by carrying some data over \service probes. 
Furthermore, one can envision enriching other congestion control protocols using \service service. We leave these as future work. 

\bibliographystyle{abbrv}
\bibliography{bibfile}

\appendix
\FloatBarrier 

\section{The analysis of \gls{dwtcp}'s theoretical properties}\label{SEC:analysis}

To achieve low latency at the queues within the network, \proto controls the congestion windows of sources to keep the utilization of resources close to the maximum link capacity. Such behavior is fulfilled as follows:

\begin{equation*}
    w \gets
    \begin{cases}
            w + L + & d_s \le d_{t} \\
            w \cdot (1 - \beta \sqrt{w} \frac{(d_s - d_{t})}{d_s}) & d_s > d_{t}
    \end{cases}
\end{equation*}

Using fluid model, we can formulate this equation as a differential equation as follows:
\begin{equation}
    w' \gets \begin{cases}
        (L + S) / \tau & t \le d_{t} \\
        - \frac{w \cdot \sqrt{w} \cdot \beta }{\tau} \cdot \frac{(d_s - d_{t})}{d_s} & d_s > d_{t}
    \end{cases}
    \label{EQ:dw}
\end{equation}

$d_s$ and $d_{t}$ can be defined as:
\begin{equation}
\label{EQ:ds_dt}
\begin{split}
    d_{s} &= \tau + \bar{k} \frac{L}{C} \\
    d_{t} &= \tau + k \frac{L}{C} \\
\end{split}
\end{equation}
Where $\bar{k} > k$.

For the sake of simplicity, let's redefine the decreasing part of the equation to become:

\begin{equation}
    w' \gets \begin{cases}
        (L + S) / \tau & t \le d_{t} \\
        - \frac{w \cdot \beta \cdot (d_s - d_{t})}{\tau \cdot d_t} & d_s > d_{t}
    \end{cases}
    \label{EQ:dw}
\end{equation}

One can notice that we removed $\sqrt{w}$ as it is originally used to enhance the fairness of the system. In addition, we replaced $d_s$ in the denominator by $d_t$ to simplify the equation. Because $d_s > d_t$ in the decrease phase, using a lower value in the denominator forces the system to be more aggressive in the decrease. Hence, if the system is stable in such a state, it is expected to be also stable if the changes are less aggressive. \\

To study the stability of such a system we tackle the two phases separately. First, the decreasing phase defined as:

\begin{equation*}
    w' = - \frac{w \cdot \beta}{\tau} \frac{(d_s - d_{t})}{d_t}
\end{equation*}

By substituting $d_s$ and $d_t$ using (\ref{EQ:ds_dt}), we get

\begin{equation*}
\begin{split}
    w' &= - \frac{w \cdot \beta}{\tau} \frac{(d_s - d_{t})}{d_t}\\
    &= - \frac{w \cdot \beta}{\tau} \frac{(\tau + \bar{k} \frac{L}{C} - \tau - k \frac{L}{C})}{\tau + k \frac{L}{C}} \\
    &= - \frac{w \cdot \beta}{\tau} \frac{L}{C} \frac{\bar{k} - k}{\tau + k \frac{L}{C}} \\
    &= - \frac{w \cdot \beta}{\tau} \frac{\bar{k} - k}{\frac{\tau \cdot C}{L} + k} \\
    w' &= - \frac{w \cdot \beta}{\tau} \frac{\bar{k} - k}{BDP + k} \\
\end{split}
\end{equation*}

Such an Equation expresses the relationship between the derivative of the congestion window $w'$ and bandwidth-delay product $BDP$ where $BDP=\frac{\tau \cdot C}{L}$.

For $k=0$; i.e., $d_{t} = \tau$ where $\tau$ is RTT, the decreasing function becomes:
\begin{equation*}
    w' = - \frac{w \cdot \beta}{\tau} \frac{\bar{k}}{BDP}
\end{equation*}

By adding the queue changing rate, one can summarize the \gls{cwnd} decreasing phase as follows:

\begin{equation}
\begin{split}
    w' & = - \beta \frac{w}{\tau} \cdot \frac{\bar{k}}{BDP} \\
    q' & = N \frac{w}{\tau} - C
\end{split}
\label{EQ:dw_decrease}
\end{equation}

On the other hand, the \gls{cwnd} increase phase can also be summarized as follows:

\begin{equation}
\begin{split}
    w' & = \frac{L+S}{\tau} \\
    q' & = N \frac{w}{\tau} - C
\end{split}
\label{EQ:dw_increase}
\end{equation}

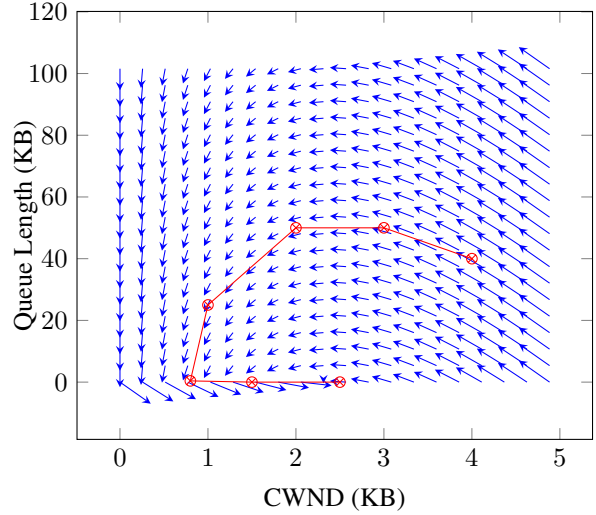
\begin{figure}[ht!]
    \centering
    \begin{tikzpicture}
        \begin{axis}[xlabel=CWND (KB), ylabel=Queue Length (KB),
            y label style={at={(0.08,0.5)}},
        ]
        \addplot[blue,quiver={u=\thisrow{u},v={20*\thisrow{v}}, scale arrows=0.35},-stealth] table[col sep=comma] {figures/stability/quiver-data.txt};
        \addplot[
            color=red,
            mark=otimes,
            ]
            coordinates {
            (4,40)(3,50)(2,50)(1,25)(.8,0.4)(1.5, 0)(2.5, 0)
        };
    \end{axis}
    \end{tikzpicture}
    \caption{Phase space trajectory of the \proto. One can notice that \proto is a stable spiral system and, it has one stable equilibrium point at ${Q=0, w=BDP/N}$.}
\label{FIG:pplane}
\end{figure}

Figure~\ref{FIG:pplane} depicts the phase trajectory of \proto dynamics after numerically solving Equations~\ref{EQ:dw_decrease} and \ref{EQ:dw_increase}. One can notice that such a system is a stable spiral system and, it has one stable equilibrium point at ${Q=0, w=BDP/N}$. For instance, if such a system starts at a state with high $w$, it reacts by reducing $w$ while increasing queue size for a while (depicted by the red path in Figure~\ref{FIG:pplane}). Thereafter, it surpasses the fair-share value of $w$ to be able to reduce queue size. Finally, it recovers by increasing $w$ again to reach the fair share.

\begin{theorem}
    \proto is spiral stable and has one stable equilibrium point as long as the number of packet processed in the high priority queue before serving any \service packets ($\bar{k}$) is less than $2 \frac{BDP}{\sqrt{\beta}}$.
\end{theorem}

\begin{proof}

Lyapunov has shown that an \gls{ode} of a nonlinear dynamical system ($x'=f(x(t))$) is said to be stable if all of its solutions start sufficiently close to the equilibrium point $x_e$ and remains close to it as $t$ increases \cite{stability_lyapunov}.
%
%
I.e., $x(t)$ is Lyapunov stable if for every $\epsilon > 0$, there exists a $\delta > 0$ such that, if $\|x(0) - x_e\| < \delta$, then for every $t \geq 0 $ we have $\|x(t) - x_e\| < \epsilon$.
More strongly, for a Lyapunov stable function $x(t)$, it is said to be asymptotically stable if $\lim_{t\rightarrow \infty}\|x(t)-x_{e}\|=0$. To that end, we prove that all solutions of \proto \gls{ode} stay close to the equilibrium point.


For the decreasing phase depicted by Equation~\ref{EQ:dw_decrease}, we deduce the solutions of this differential equation by extracting the Jacobian matrix as follows:

\begin{equation}
    J =
    \begin{pmatrix}
        \frac{\partial w'}{\partial w} & \frac{\partial w'}{\partial q}\\
        \frac{\partial q'}{\partial w} & \frac{\partial q'}{\partial q}
    \end{pmatrix} =
    \begin{pmatrix}
        - \frac{\beta \cdot \bar{k}}{\tau \cdot BDP} & -\frac{\beta \cdot w}{\tau \cdot BDP} \\
        \frac{N}{\tau} & 0
    \end{pmatrix}
\end{equation}

The eigenvalues of a $2 \times 2$ matrix are the solutions of the determinant equation
$|\lambda I - j| = 0$

\begin{equation}
    \begin{vmatrix}
        \lambda + \frac{\beta \cdot \bar{k}}{\tau \cdot BDP} & \frac{\beta \cdot w}{\tau \cdot BDP} \\
        -\frac{N}{\tau} & \lambda
    \end{vmatrix} = 0
\end{equation}

\begin{equation}
    \lambda^2 + \lambda \frac{\beta \cdot \bar{k}}{\tau \cdot BDP} + \frac{\beta \cdot w \cdot N}{\tau^2 \cdot BDP} = 0
\end{equation}

The roots of the previous equation can be calculated as follows:
\begin{equation}
\begin{split}
    r_\lambda &= \frac{-b \pm \sqrt{b^2 - 4ac}}{2a} \\
     &= \frac{-\frac{\beta \cdot \bar{k}}{\tau \cdot BDP} \pm \sqrt{(\frac{\beta \cdot \bar{k}}{\tau \cdot BDP})^2 - 4 (\frac{\beta \cdot w \cdot N}{\tau^2 \cdot BDP})}}{2}
\end{split}
\end{equation}

For a stable spiral system, the eigenvalues must be complex numbers with a negative real part. To satisfy such a requirement, Inequality (\ref{EQ:stabilityCond1}) must hold to guarantee that the eigenvalues have a negative real part. In addition, Inequality (\ref{EQ:stabilityCond2}) must hold to guarantee that eigenvalues are complex numbers.

\begin{equation}
    -\frac{\beta \cdot \bar{k}}{\tau \cdot BDP} < 0
\label{EQ:stabilityCond1}
\end{equation}

Which is always true for all values of $\beta, \bar{k}, \tau, $ and $BDP$.


\begin{equation}
\begin{split}
    (\frac{\beta \cdot \bar{k}}{\tau \cdot BDP})^2 - 4 \frac{\beta \cdot w \cdot N}{\tau^2 \cdot BDP} &< 0 \\
    \frac{\beta^2 \cdot \bar{k}^2}{\tau^2 \cdot BDP^2} - 4 \frac{\beta \cdot w \cdot N}{\tau^2 \cdot BDP} &< 0 \\
    \frac{\beta \cdot \bar{k}^2}{BDP} - 4 w \cdot N &< 0 \\
    \beta \cdot \bar{k}^2 - 4 w \cdot N \cdot BDP &< 0 \\
    \beta \cdot \bar{k}^2 &< 4 w \cdot N \cdot BDP
\end{split}
\label{EQ:stabilityCond2}
\end{equation}

At the equilibrium point, $BDP = w \cdot N$. Therefore, we can reduce Equation~(\ref{EQ:stabilityCond2}) to become

\begin{equation}
    \begin{split}
        \beta \cdot \bar{k}^2 &< 4 BDP^2 \\
        \bar{k} &< 2 \frac{BDP}{\sqrt{\beta}}
    \end{split}
    \label{EQ:stabilityCond2_2}
\end{equation}

Hence, one can conclude that \gls{dwtcp} decreasing phase is stable as long as the number of processed packet in the high priority queue before serving any \service $(\bar{k})$ is less than $2 \frac{BDP}{\sqrt{\beta}}$; i.e., Inequality (\ref{EQ:stabilityCond2_2}) must be satisfied. \\

For the increasing phase stated in Equation~(\ref{EQ:dw_increase}), we can represent Jacobian matrix as follows:

\begin{equation}
    J =
    \begin{pmatrix}
        0 &0\\
        0 & \frac{N}{\tau}
    \end{pmatrix}
\end{equation}

The eigenvalues of this matrix are the solutions of the determinant equation $|\lambda I - j| = 0$.

\begin{equation}
\begin{split}
    \begin{vmatrix}
        \lambda & 0\\
        0 & \lambda-\frac{N}{\tau}
    \end{vmatrix} = 0 \\
    \lambda^2 -\frac{N}{\tau} \lambda = 0 \\
    (\lambda = 0) \text{ and } (\lambda = \frac{N}{\tau})
\end{split}
\end{equation}

Thus, the increasing phase is not stable for \gls{dwtcp}. However, the decreasing phase is triggered when $d_s > \tau$. Hence, we conclude that the stability of \gls{dwtcp} is controlled by the decreasing phase.

\end{proof}

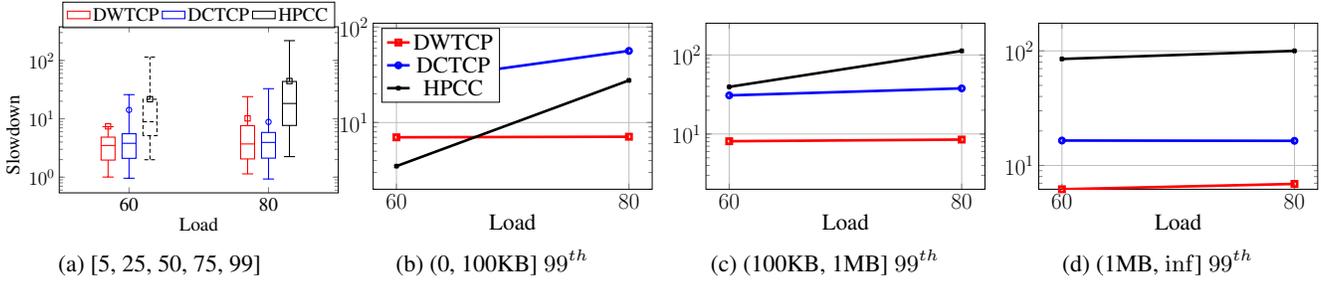
\begin{figure*}[htbp]
    \centering
    \begin{subfigure}[b]{0.24\textwidth}
        \centering

       \begin{tikzpicture}[scale=0.5, font=\Large]
        \begin{axis} [area legend,legend columns=-1, legend style={at={(0.5,1.15)},anchor=north,legend cell align=left, fill opacity=0.6, text opacity=1},
        height=6.0cm, width=9cm,
        xmin=0, xmax=25, 
          xtick={1,3},
          xticklabels={60, 80},
          xmin=0,xmax=4, xlabel=Load,
          ylabel= Slowdown,  y label style={at={(0.0,0.35)}},
          ymode=log, boxplot/draw direction=y,
          /pgfplots/boxplot/box extend=0.2
          ]

          
        \addplot+ [color=red,thick,mark=square,
        boxplot prepared={
            median=3.70650,
            upper quartile=7.63127,
            lower quartile=2.05356,
            upper whisker=23.9635,
            lower whisker=1.138315,
            draw position=2.7
        },
        ] coordinates {(2.7,10.16410)};
        \addlegendentry{DWTCP}

        \addplot+ [color=blue,thick, mark=o,
          boxplot prepared={
            median=3.93889,
            upper quartile=5.8353,
            lower quartile=2.11810,
            upper whisker=32.81679,
            lower whisker=0.9296026,
            draw position=3.0
          },
          ] coordinates {(3.0,8.83840)};
          \addlegendentry{DCTCP}

          \addplot+ [color=black,thick,mark=square,
          boxplot prepared={
            median=18.3,
            upper quartile=43.590517,
            lower quartile=7.69,
            upper whisker=218.13233,
            lower whisker=2.249,
            draw position=3.3
          },
          ] coordinates {(3.3,44.33)};
          \addlegendentry{HPCC}

          \addplot+ [color=blue, thick, mark=o,
          boxplot prepared={
            median=3.80321,
            upper quartile=5.54611,
            lower quartile=2.10227,
            upper whisker=25.94041,
            lower whisker=0.95835,
            draw position=1.0
          },
          ] coordinates {(1.0,14.1976)};

          \addplot+ [color=red,mark=square, thick, 
          boxplot prepared={
            median=3.494497,
            upper quartile=4.85102,
            lower quartile=1.961481,
            upper whisker=7.31557,
            lower whisker=1.004320,
            draw position=0.7
          },
          ] coordinates {(0.7,7.429018)};
          
          \addplot+ [color=black,thick,mark=square,
          boxplot prepared={
            median=8.9,
            upper quartile=21.590517,
            lower quartile=5.14,
            upper whisker=114.13233,
            lower whisker=1.99,
            draw position=1.3
          },
          ] coordinates {(1.3,21.66)};

        \end{axis}
      \end{tikzpicture}
        \caption{[5, 25, 50, 75, 99]} 
    \end{subfigure}
    \hspace*{\fill}
    \begin{subfigure}[b]{0.24\textwidth}
        \centering
        \begin{tikzpicture}[scale=0.5, font=\LARGE]
            \begin{axis}[
                ymin=2,
                width=9cm,
                height=6cm,
                xlabel=Load,
                ymode=log,
                symbolic x coords={$20$, $60$, $80$},
                xtick = data,
                enlarge y limits={value=0.2,upper},
                legend pos=north west,
                ymajorgrids=true,xmajorgrids=true,
            ]


            \addplot[color=red, mark=square, line width=2.0pt] coordinates {($60$, 7) ($80$, 7.1)};
            \addplot[color=blue, mark=o, line width=2.0pt] coordinates { ($60$, 26) ($80$, 56.3)};
            \addplot[color=black, mark=x, line width=2.0pt] coordinates {($60$, 3.49) ($80$, 27.64)};
            \legend{DWTCP, DCTCP, HPCC}
            \end{axis}
        \end{tikzpicture}
        \caption{(0, 100KB] $99^{th}$} 
        \label{FIG:websearch_small}
    \end{subfigure}
    \hspace*{\fill}
    \begin{subfigure}[b]{0.24\textwidth}
        \centering
        \begin{tikzpicture}[scale=0.5, font=\LARGE]
            \begin{axis}[
                ymin=2,
                width=9cm,
                height=6cm,
                xlabel=Load,
                ymode=log,
                symbolic x coords={$20$, $60$, $80$},
                xtick = data,
                enlarge y limits={value=0.2,upper},
                legend pos=north west,
                ymajorgrids=true,xmajorgrids=true,
            ]
            \addplot[color=red, mark=square, line width=2.0pt] coordinates {($60$, 8.1) ($80$, 8.5)};
            \addplot[color=blue, mark=o, line width=2.0pt] coordinates {($60$, 30.8) ($80$, 37.8)};
            \addplot[color=black, mark=x, line width=2.0pt] coordinates {($60$, 39.49) ($80$, 112.64)};
            \end{axis}
        \end{tikzpicture}
        \caption{(100KB, 1MB] $99^{th}$} 

    \end{subfigure}
    \hspace*{\fill}
    \begin{subfigure}[b]{0.24\textwidth}
        \centering
        \begin{tikzpicture}[scale=0.5, font=\LARGE]
            \begin{axis}[
                ymin=0,
                width=9cm,
                height=6cm,
                xlabel=Load,
                ymode=log,
                symbolic x coords={$20$, $60$, $80$},
                xtick = data,
                enlarge y limits={value=0.2,upper},
                legend pos=north west,
                ymajorgrids=true,xmajorgrids=true,
            ]

            \addplot[color=red, mark=square, line width=2.0pt] coordinates {($60$, 6.2) ($80$, 6.9)};
            \addplot[color=blue, mark=o, line width=2.0pt] coordinates {($60$,  16.5) ($80$, 16.4)};
            \addplot[color=black, mark=x, line width=2.0pt] coordinates {($60$, 85) ($80$, 100)};
            \end{axis}
        \end{tikzpicture}
        \caption{(1MB, $\inf$] $99^{th}$} 
    \end{subfigure}

        \caption{Websearch workload: a) Percentile plots. Small circle and square show the mean slowdown. Compared to DCTCP (and HPCC)\proto improves up to 2x (3x) mean slowdown and up to 3.5x (5x) for the tail slowdown. (b-d) \proto significantly improves tail slowdown for small by 8x (3.7x), medium by 3.5x (10x) and large flows by 3.5x (15x).}
        
        \label{FIG:websearch}
\end{figure*}

\section{Extra Results}
\label{sec:results}

Here we provide additional results for understanding \proto performance.

\paragraph{\textbf{Websearch workload}}

Here, we present the slowdown in flow completion time of the websearch workload in a leaf-spine topology as explained in~\S\ref{sec:benchmark}.
As shown in Figure~\ref{FIG:websearch}, \proto reduces the mean slowdown by 2x and the $99^{th}$ percentile by 3.5x (for the 60\% load).
Also, the improvement in the tail slowdown is much more significant for smaller flows (up to 8x as shown in Figure~\ref{FIG:websearch_small}).

\begin{figure*}[htbp]
    \centering
    \begin{subfigure}[b]{0.32\linewidth}
        \centering
        \begin{tikzpicture}[scale=0.5,font=\Large]
            \begin{axis}[legend style={at={(0.5,1.15)}, legend columns=-1, anchor=north,legend cell align=left, fill opacity=0.6, text opacity=1},
                xlabel=Time (s), ylabel=Rate (Mbps), y label style={at={(0.0,0.5)}}, height=6.cm, width=9cm,
                ymajorgrids=true,xmajorgrids=true,enlarge y limits={value=0.2,upper},]
                \addplot[color=olive, mark=diamond, line width=2.0pt, each nth point=2] table [x=x, y=y, col sep=comma] {figures/varying_sendersnum/30senders/goodput_mbps-sum.csv};
                \addlegendentry{30 senders}
                \addplot[color=darkgray, mark=o, line width=2.0pt, each nth point=2] table [x=x, y=y, col sep=comma] {figures/varying_sendersnum/20senders/goodput_mbps-sum.csv};
                \addlegendentry{20 senders}
                \addplot[color=magenta, mark=x, line width=2.0pt, each nth point=2] table [x=x, y=y, col sep=comma] {figures/varying_sendersnum/10senders/goodput_mbps-sum.csv};
                \addlegendentry{10 senders}
            \end{axis}
        \end{tikzpicture}
        \caption{Throughput}
        \label{FIG:varyinge_senders_goodput}
    \end{subfigure}
    \hspace*{\fill}
    \begin{subfigure}[b]{0.32\linewidth}
        \centering
        \begin{tikzpicture}[scale=0.5,font=\Large]
            \begin{axis}[legend style={at={(0.5,1.15)}, legend columns=-1,,anchor=north,legend cell align=left, fill opacity=0.6, text opacity=1},
                xlabel=Time (s), ylabel=Queue Length (Packet), y label style={at={(0.0,0.5)}},
                , height=6.cm, width=9cm,
                ymajorgrids=true,xmajorgrids=true,]
                \addplot[color=olive, mark=diamond, line width=2.0pt, each nth point=3] table [x=x, y=y, col sep=comma] {figures/varying_sendersnum/30senders/packetsInQueue-r0-q0.csv};
                \addlegendentry{30 senders}
                \addplot[color=darkgray, mark=o, line width=2.0pt, each nth point=3] table [x=x, y=y, col sep=comma] {figures/varying_sendersnum/20senders/packetsInQueue-r0-q0.csv};
                \addlegendentry{20 senders}
                \addplot[color=magenta, mark=x, line width=2.0pt, each nth point=3] table [x=x, y=y, col sep=comma] {figures/varying_sendersnum/10senders/packetsInQueue-r0-q0.csv};
                \addlegendentry{10 senders}
            \end{axis}
        \end{tikzpicture}
        \caption{Queue Length}
        \label{FIG:varying_senders_num_queuesize}
    \end{subfigure}
    \hspace*{\fill}
    \begin{subfigure}[b]{0.32\linewidth}
        \centering
        \begin{tikzpicture}[scale=0.5,font=\Large]
            \begin{axis}[legend style={at={(0.5,1.15)}, legend columns=-1, anchor=north,legend cell align=left, fill opacity=0.6, text opacity=1},
                xlabel=Time (s), ylabel=Jain Fairness Index, y label style={at={(0.0,0.5)}}, height=6.cm, width=9cm,
                ymajorgrids=true,xmajorgrids=true,enlarge y limits={value=0.2,upper},]
                \addplot[color=olive, mark=diamond, line width=2.0pt, each nth point=2] table [x=x, y=y, col sep=comma] {figures/varying_sendersnum/30senders/fairness-index.csv};
                \addlegendentry{30 senders}
                \addplot[color=darkgray, mark=o, line width=2.0pt, each nth point=2] table [x=x, y=y, col sep=comma] {figures/varying_sendersnum/20senders/fairness-index.csv};
                \addlegendentry{20 senders}
                \addplot[color=magenta, mark=x, line width=2.0pt, each nth point=2] table [x=x, y=y, col sep=comma] {figures/varying_sendersnum/10senders/fairness-index.csv};
                \addlegendentry{10 senders}
            \end{axis}
        \end{tikzpicture}
        \caption{Fairness Index}
        \label{FIG:varyinge_senders_num_fairness}
    \end{subfigure}

    \caption{Varying the number of senders. \proto maintains high throughput, low queue size, and fairness for epxeriments with different number of senders.}
    \label{FIG:varyinge_senders_num}
\end{figure*}
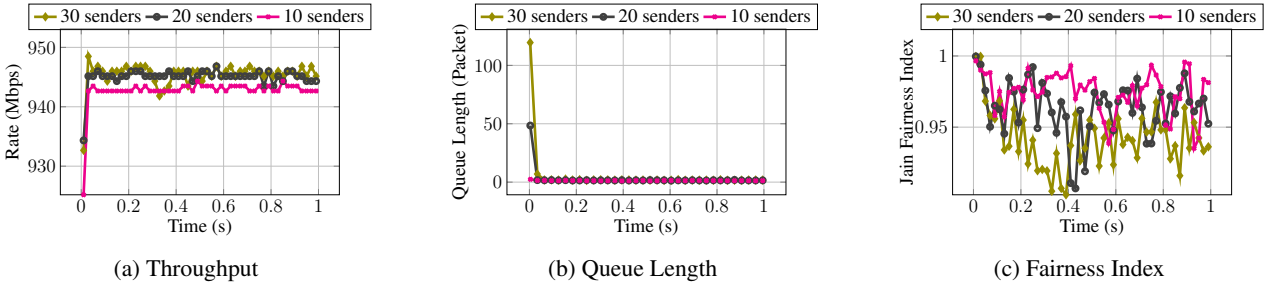
\paragraph{\textbf{Effect of varying number of senders}}
In this experiment, we set the number of senders (receivers) to 10, 20, and 30 and we the default \proto parameters.
We can verify that \proto adapts to different number of flows and maintains small queues (Figure~\ref{FIG:varying_senders_num_queuesize}). It also achieves high throughput and better fairness as depicted in Figures~\ref{FIG:varyinge_senders_goodput} and \ref{FIG:varyinge_senders_num_fairness} respectively.
We observe an increase in the queue size in the beginning, which is the result of synchronization among the flows.

\begin{figure*}[ht!]
    \centering
    \begin{subfigure}[b]{0.32\linewidth}
        \centering
        \begin{tikzpicture}[scale=0.6,font=\Large]
            \begin{axis}[legend style={at={(0.2,.95)},anchor=north,legend cell align=left, fill opacity=0.6, text opacity=1},
                xlabel=Time (s), ylabel=Rate (Gbps), y label style={at={(0.08,0.5)}}, height=6.cm, width=8cm,
                ymajorgrids=true,xmajorgrids=true,]
                \addplot[color=black, mark=pentagon, line width=2.0pt, each nth point=5] table [x=x, y=y, col sep=comma] {figures/varying_bw/newreno/dequeue_rate-r0_0.csv};
                \addlegendentry{TCP}
                \addplot[color=blue, mark=square, line width=2.0pt, each nth point=5]  table [x=x, y=y, col sep=comma] {figures/varying_bw/dctcp/dequeue_rate-r0_0.csv};
                \addlegendentry{DCTCP}
                \addplot[color=red, mark=+, line width=2.0pt, each nth point=5]  table [x=x, y=y, col sep=comma] {figures/varying_bw/dwtcp/dequeue_rate-r0_0.csv};
                \addlegendentry{DWTCP}
            \end{axis}
        \end{tikzpicture}
        \caption{Throughput}
        \label{FIG:varybw1}
    \end{subfigure}
    \hspace*{\fill}
    \begin{subfigure}[b]{0.32\linewidth}
        \centering
        \begin{tikzpicture}[scale=0.6,font=\Large]
            \begin{axis}[legend style={at={(0.2,.95)},anchor=north,legend cell align=left, fill opacity=0.6, text opacity=1},
                xlabel=Time (s), ylabel=Queue Length (Packet), y label style={at={(0.03,0.5)}}, height=6.cm, width=8cm,
                ymajorgrids=true,xmajorgrids=true,]
                \addplot[color=black, mark=pentagon, line width=2.0pt, each nth point=5]  table [x=x, y=y, col sep=comma] {figures/varying_bw/newreno/packetsInQueue-r0-q0.csv};
                \addlegendentry{TCP}
                \addplot[color=blue, mark=square, line width=2.0pt, each nth point=5]  table [x=x, y=y, col sep=comma] {figures/varying_bw/dctcp/packetsInQueue-r0-q0.csv};
                \addlegendentry{DCTCP}
                \addplot[color=red, mark=+, line width=2.0pt, each nth point=5]  table [x=x, y=y, col sep=comma] {figures/varying_bw/dwtcp/packetsInQueue-r0-q0.csv};
                \addlegendentry{DWTCP}
            \end{axis}
        \end{tikzpicture}
        \caption{Queue Length}
        \label{FIG:varybw2}
    \end{subfigure}
    \hspace*{\fill}
    \begin{subfigure}[b]{0.32\linewidth}
        \centering
        \begin{tikzpicture}[scale=0.6,font=\Large]
            \begin{axis}[legend style={at={(0.2,.5)},anchor=north,legend cell align=left, fill opacity=0.6, text opacity=1},
                xlabel=Time (s), ylabel=Jain Fairness Index, y label style={at={(0.03,0.5)}}, height=6.cm, width=8cm,
                ymajorgrids=true,xmajorgrids=true,]
                \addplot[color=black, mark=pentagon, line width=2.0pt, each nth point=3] table [x=x, y=y, col sep=comma] {figures/varying_bw/newreno/fairness-index.csv};
                \addlegendentry{TCP}
                \addplot[color=blue, mark=square, line width=2.0pt, each nth point=3]  table [x=x, y=y, col sep=comma] {figures/varying_bw/dctcp/fairness-index.csv};
                \addlegendentry{DCTCP}
                \addplot[color=red, mark=+, line width=2.0pt, each nth point=3] table [x=x, y=y, col sep=comma] {figures/varying_bw/dwtcp/fairness-index.csv};
                \addlegendentry{DWTCP}
            \end{axis}
        \end{tikzpicture}
        \caption{Fairness}
        \label{FIG:varybw3}
    \end{subfigure}
    \caption{Varying the link bandwidth. \proto adpats to bandwidth increases and decreases very quickly without any queue buildups or sacrificing fairness.}
    \label{FIG:varying_bw}
\end{figure*}
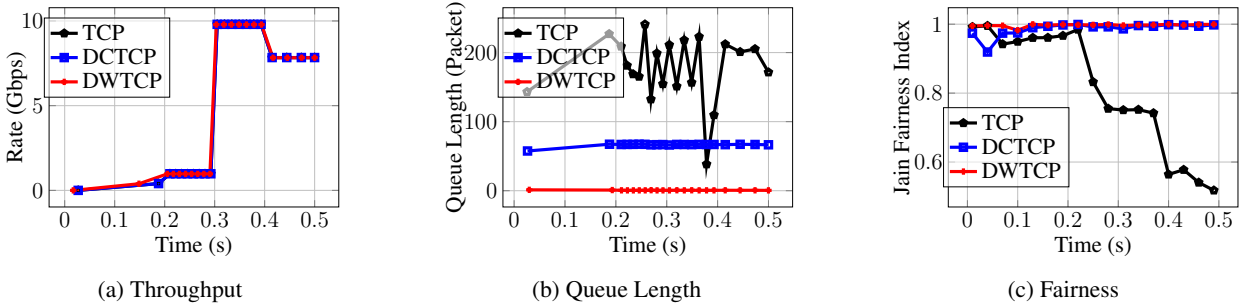

\paragraph{\textbf{Effect of varying network conditions}}
To observe the adaptability of \service to varying network conditions, we vary the link bandwidth every 0.1 seconds as [0.4, 0.7, 1, 10, 0.8]~Gbps.
We run five long flows to observe the queue size, throughput, and fairness.
We observe that \proto quickly grabs any available bandwidth when the available bandwidth changes from 1Gbps to 10Gbps and it quickly converges to full link utilization as shown in Figure~\ref{FIG:varybw1}.
%
There are two important observations: First, as the link capacity grows and there is less room, \proto adapts its aggressiveness to avoid any queue buildups (Figure~\ref{FIG:varybw2}). \service incurs very little overhead and \service rate adjusts with the variation in link rate ensuring fairness among the flows (Figure~\ref{FIG:varybw3}).
Second, when the available bandwidth decreases, \service detects it quickly and flows give up the extra bandwidth without causing any queue buildups and losing on fairness as is the case with TCP.

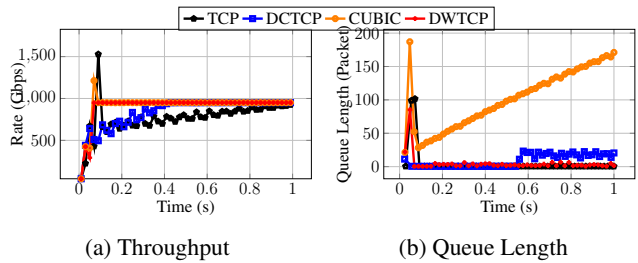
\begin{figure}[ht!]
    \centering
    \begin{subfigure}[b]{0.48\linewidth}
        \centering
        \begin{tikzpicture}[scale=0.45,font=\Large]
            \begin{axis}[legend style={at={(1.1,1.15)}, legend columns=-1, anchor=north,legend cell align=left, fill opacity=0.6, text opacity=1},
                xlabel=Time (s), ylabel=Rate (Gbps), y label style={at={(0.0,0.5)}}, height=6.cm, width=9cm,
                ymajorgrids=true,xmajorgrids=true,enlarge y limits={value=0.2,upper},]
                \addplot[color=black, mark=pentagon, line width=2.0pt, each nth point=2] table [x=x, y=y, col sep=comma] {figures/longhaul/tcp/goodput_mbps-sum.csv};
                \addlegendentry{TCP}
                \addplot[color=blue, mark=square, line width=2.0pt, each nth point=2] table [x=x, y=y, col sep=comma] {figures/longhaul/dctcp/goodput_mbps-sum.csv};
                \addlegendentry{DCTCP}
                 \addplot[color=orange, mark=o, line width=2.0pt, each nth point=2] table [x=x, y=y, col sep=comma] {figures/longhaul/cubic/goodput_mbps-sum.csv};
                \addlegendentry{CUBIC}
                \addplot[color=red, mark=+, line width=2.0pt, each nth point=2] table [x=x, y=y, col sep=comma] {figures/longhaul/dwtcp/goodput_mbps-sum.csv};
                \addlegendentry{DWTCP}
            \end{axis}
        \end{tikzpicture}
        \caption{Throughput}
        \label{FIG:longhaulIncreasing1}
    \end{subfigure}
    \hspace*{\fill}
    \begin{subfigure}[b]{0.48\linewidth}
        \centering
        \begin{tikzpicture}[scale=0.45,font=\Large]
            \begin{axis}[
                xlabel=Time (s), ylabel=Queue Length (Packet), y label style={at={(0.01,0.5)}},
                , height=6.cm, width=9cm,
                ymajorgrids=true,xmajorgrids=true,]
                \addplot[color=black, mark=pentagon, line width=2.0pt, each nth point=2] table [x=x, y=y, col sep=comma] {figures/longhaul/tcp/packetsInQueue-r0-q0.csv};
                \addlegendentry{TCP}
                \addplot[color=blue, mark=square, line width=2.0pt, each nth point=2] table [x=x, y=y, col sep=comma] {figures/longhaul/dctcp/packetsInQueue-r0-q0.csv};
                \addlegendentry{DCTCP}
                \addplot[color=orange, mark=o, line width=2.0pt, each nth point=2] table [x=x, y=y, col sep=comma] {figures/longhaul/cubic/packetsInQueue-r0-q0.csv};
                \addlegendentry{CUBIC}
                \addplot[color=red, mark=+, line width=2.0pt, each nth point=2] table [x=x, y=y, col sep=comma] {figures/longhaul/dwtcp/packetsInQueue-r0-q0.csv};
                \addlegendentry{DWTCP}
                \legend{};
            \end{axis}
        \end{tikzpicture}
        \caption{Queue Length}
        \label{FIG:longhaulIncreasing2}
    \end{subfigure}
    \caption{Long haul network experiment. \proto converges to the optimal throughput in high BDP faster than other protocols with no queue buildups.}
    \label{FIG:longhaulIncreasing}
\end{figure}

\paragraph{\textbf{Performance in large BDP networks.}}
In this experiment, we evaluate the performance of \gls{dwtcp} for long haul 
such as intra-data center networks.
Here, the challenge is that the \gls{bdp} is very high, and the buffer sizes are much lower than the \gls{bdp}. Moreover, the \gls{tcp} feedback loop is very long, resulting in a very long convergence time for \gls{tcp}'s congestion window to reach its fair share of the high \gls{bdp}. In this experiment, We generate 5 long flows, set the link capacity to 1Gbps and set RTT to $6ms$.

Figure~\ref{FIG:longhaulIncreasing1} shows the throughput of \proto, TCP and DCTCP over time. As shown, TCP takes around 1~second and DCTCP around 0.4~seconds to converge to the optimal throughput due to the high \gls{bdp}.
On the other hand, the congestion window in \proto reaches its optimal value quickly showing orders of magnitude faster convergence time as compared to DCTCP and TCP.
Also, note that the \service rate at the switch adapts quickly based on the available bandwidth and \proto converges quickly without any queue buildups as shown in Figure~\ref{FIG:longhaulIncreasing2} which is not the case for CUBIC.

\end{document}